\documentclass[preprint,showpacs,preprintnumbers,amsmath,amssymb]
{revtex4}

\usepackage{graphicx}
\usepackage{dcolumn}
\usepackage{bm}

\begin{document}

\preprint{Typeset by REV\TeX~4}

\title{Eigenmodes and growth rates of relativistic
current filamentation instability in a collisional plasma}

\author{M.~Honda}
\affiliation{Plasma Astrophysics Laboratory, Institute for Global Science,
Mie 519-5203, Japan}

\begin{abstract}
I theoretically found eigenmodes and growth rates
of relativistic current filamentation instability
in collisional regimes, deriving a generalized dispersion relation from
self-consistent beam-Maxwell equations.
For symmetrically counterstreaming, fully relativistic electron currents,
the collisional coupling between electrons and ions
creates the unstable modes of growing oscillation and wave,
which stand out for long-wavelength perturbations.
In the stronger collisional regime,
the growing oscillatory mode tends to be dominant for all wavelengths.
In the collisionless limit, those modes vanish,
while maintaining another purely growing mode that exactly coincides with
a standard relativistic Weibel mode.
It is also shown that
the effects of electron-electron collisions and thermal spread
lower the growth rate of the relativistic Weibel instability.
The present mechanisms of filamentation dynamics are essential for
transport of homogeneous electron beam produced by the interaction of
high power laser pulses with plasma.

\end{abstract}

\pacs{52.25.Fi, 52.27.Ny, 52.35.Qz}

\maketitle

\section{\label{sec:1}Introduction}

Relativistic laser-plasma interactions have been a topical issue
for the past decade \cite{umstadter03},
in the context of ignitor physics of
inertial confinement fusion, aiming at additional fast heating and
subsequent ignition of highly compressed targets
by means of an external intense laser pulse \cite{key02}.
The laser pulse drives relativistic currents, compensating return currents,
and creates a pattern of counterpropagating currents
which are subject to current filamentation instabilities (CFI)
including the Weibel mode \cite{weibel59}.
In the theoretical arena,
numerous versions of analytical and numerical methods have been developed
in the past, to explore this type of instabilities
\cite{bornatici70,davidson72,benford73,lee73,molvig75,lemons80,
okada80,cary81,shukla82,lee83,uhm83,hughes86,yoon87,wallace87}.
In general, the physical mechanism of the electromagnetic CFI
is explained as follows:
When the compensation of the counterpropagating electron currents is
disturbed in the transverse direction,
magnetic repulsion between the two currents reinforces the
initial disturbance.
As a consequence, a larger and larger magnetic field is produced
as time increases, degrading the transport properties.
Many efforts have been devoted to this crucial problem,
both related to laboratory electron beams
\cite{kapetanakos74,segalov80,fisher88}
as well as to astrophysics \cite{yang93,kazimura98,medvedev99}.
Concerning laser interaction with plasmas,
for the low (nonrelativistic) intensity regimes the
inverse bremsstrahlung absorption is known to predominate
around the cutoff region, where the Weibel-type instability
associated with temperature anisotropy can take place
\cite{romanov97,bendib97}.
In the ablative plasmas, the collisional and related nonlocal effects
were investigated \cite{epperline87}.

The original motivation for this work was triggered by
more recent publications that have been quantitatively treated
with the counterstreaming relativistic CFI in the collisionless limit
\cite{califano97,califano98a,califano98b}.
A series of works could be linked with the ignitor physics by
irradiating a relativistic laser pulse:
Fast ignition of the compressed fuel requires at least
$10-100~{\rm kJ}$ of external energy to be deposited within
$\sim 10~{\rm ps}$ into the precompressed core \cite{tabak94}.
If carried by $1-10~{\rm MeV}$ electrons,
it implies a current of $0.1-1~{\rm GA}$ which exceeds the
transport limit of about $100~{\rm kA}$ \cite{honda00a},
by more than a factor $\sim 10^3$.
The essential feature is the breakup of the relativistic electron beam
into many filaments when propagating in dense plasma
\cite{gremillet02,tatarakis03}.
The physics underlying this phenomenon is just the CFI as mentioned above.
This type of instability leads to nonlinear filamentation and
coalescence of the relativistic electron beam
\cite{lee73,montgomery79,pukhov97,honda00b,honda00c,kazimura01}
and the formation of strong magnetic fields
\cite{pukhov97,honda00b,honda00c,askaryan94,askaryan97}.
In the astronomical point of view,
one often encounters such morphology in a variety of celestial objects,
particularly, in the astrophysical jets \cite{honda02}.
State-of-the-art observations by utilizing very long baseline interferometry
have revealed the filamentary structure of the jets \cite{asada00},
involving transverse magnetic fields \cite{gabuzda99}.
More recently, large-scale toroidal magnetic fields have been
discovered in the Galactic Center \cite{novak03},
accompanied with splendid filamentary radio arcs
\cite{yusefzadeh84,yusefzadeh87}.

Regarding the ignitor physics of laboratory plasmas,
the effects of collisions and beam thermal spread
play a significant role in the CFI
caused by the ultrahigh relativistic currents through ablative coronal plasma
\cite{honda00c},
whose density rises from $n_c \approx 10^{21}~{\rm cm^{-3}}$ (cutoff density)
to $\sim 10^{26}~{\rm cm^{-3}}$ (a thousand times solid density),
over a radial distance of $\sim 10^2~{\rm {\mu m}}$ \cite{honda03a}.
The role of collisions in a laser-produced plasma has been
discussed in the early literature by Motz \cite{motz79}, though
the description was restricted to the nonrelativistic fashion.
It is important to note that the collision frequency
invokes additional parameter disturbing
the universal density scaling, in terms of the time scale of
a plasma oscillation period $\omega_{pe}^{-1}$
and the spatioscale of a skin depth $c/\omega_{pe}$,
which is valid only for collisionless regimes.
In fully relativistic regimes, the collision cross section should be evaluated
by using the Mott scattering formula \cite{mott61}.
Presuming a small angle scattering and averaging over the angle
yield the electron-ion collision frequency that can be defined as
$\nu_{ei} = (n_i Y_{ei}/c^3)(\Gamma/\mu^3)$,
where $n_i$ is the number density of ions,
$Y_{ei}=4 \pi (\bar{Z} e^2/m_0)^2 {\rm ln} \Lambda$,
$\Gamma = (1 + \mu^2)^{1/2}$, $\mu=p_e/m_0c$, and $p_e$, $m_0$, $\bar{Z}$,
and ${\rm ln} \Lambda$ are the electron momentum, the electron rest mass,
the averaged charge number,
and the Coulomb logarithm, respectively \cite{honda03b},
and the electron-electron collision frequency is given as
$\nu_{ee} = (2/\bar{Z}) \nu_{ei}$.
Introducing the current neutral condition of $n_b/n_p \simeq |v_p/c| \ll 1$,
where $n_b$, $n_p$, and $v_p$ are the beam electron density,
the plasma electron density, and its velocity, respectively,
we find the ratio of
$\nu_{ei}/\omega_{pe} \simeq 2 \pi^{1/2} \bar{Z}^{9/2}
e^3 n_i^{1/2} (m_0 c^2)^{-3/2} (n_i/n_b)^3 {\rm ln} \Lambda$
for the plasma return current.
When assuming $n_b \sim n_c \approx 10^{21}~{\rm cm}^{-3}$
to be almost constant, the ratio can be estimated as

\begin{widetext}
\begin{equation}
{\nu_{ei} \over \omega_{pe}} \sim \bar{Z}^{9/2}
\left( n_i \over {10^{23}~{\rm cm}^{-3}} \right)^{7/2}
\left( { {\rm ln} \Lambda \over 5} \right).
\label{eq:0}
\end{equation}
\end{widetext}

\noindent
Equation~(\ref{eq:0}) indicates that indeed,
collisional effects are important in the supersolid density regions of
$n_i \geq 10^{23}~{\rm cm}^{-3}$, and particularly,
for the Lorentz plasma with $\bar{Z} \gg 1$.
In addition, the beam thermal spread will occur there,
since the electrons penetrating through the ablative corona
are expected to be thermalized via the collisional and
collisionless dissipative processes \cite{honda00b}.
The thermal effects involved with the ratio of the
transverse temperature to the total energy of beam electron,
i.e., $\sim T_{b \perp}/(\Gamma m_0 c^2)$
with expected values in the range less than unity \cite{honda00c},
which also disturbs the aforementioned scaling.

In this paper, I present fundamental eigenmode properties and growth rates in
the linear stage of the CFI including the effects that violate
the universal density scaling.
The theory is expanded in fully relativistic regime
on the basis of the self-consistent beam-Maxwell equations.
To the best of my knowledge, a comprehensive treatment of
electron-electron and electron-ion collisional effects
on relativistic counterstreaming electron currents with thermal spread
has not been carried out, so far.
For the symmetrically counterstreaming currents,
I found that collisional and thermal effects
are likely to lower the growth rate of the relativistic Weibel instability.
The most significant result is that
the finite collisional coupling between electron and ion creates
the growing oscillatory and the growing wave modes, which stand out
for long-wavelength perturbations,
and in the moderate to strong collisional regime,
even for short-wavelength perturbations,
the growth rate of the oscillatory mode exceeds that of the
suppressed Weibel instability.
In this aspect, the present work goes beyond the framework
of the well-established theory of electromagnetic instabilities.
The asymmetric configuration effects of the counterstreaming currents
are also investigated by using a slow return current approximation.
Furthermore, I argue that the collisional coupling
between electron and electron creates a growing wave mode,
but its growth rate is lower than that of the suppressed Weibel instability.
It is also shown that thermal effects participate in lowering
the growth rate of the Weibel instability.
Although, in the case that includes thermal corrections,
the present calculation is valid for the smaller wave number $k$
as shown later, the most interesting range
$k \sim \omega_{pe}/c$ can be fairly covered.

In order to spell out these subjects,
the present paper is organized as follows:
In Sec.~\ref{sec:2}, the linear theoretical analysis of the
relativistic CFI is expanded systematically.
The basic equations introduced in Sec.~\ref{sec:2a}
are linearized so as to obtain a dispersion relation of the CFI
along the manner outlined in the Appendix.
The generic dispersion relation is presented in Sec.~\ref{sec:2b}, and
its approximate expressions are derived in Sec.~\ref{sec:2c}.
In Sec.~\ref{sec:3}, for an application,
the newly derived equations are solved for typical parameters of
counterstreaming relativistic currents,
and the properties of complex eigenmodes are investigated
for the cases including the effects of
electron-electron collision (Sec.~\ref{sec:3a}),
electron-ion collision (Sec.~\ref{sec:3b}),
thermal corrections (Sec.~\ref{sec:3c}), and for
the special case without collisional and thermal effects (Sec.~\ref{sec:3d}),
as well as, for the case with an asymmetrical configuration of
counterstreaming currents (Sec.~\ref{sec:3e}),
closely relevant to ignitor physics.
As a matter of convenience, formulas of the growth rates
are explicitly written down for some interesting cases.
Finally, Sec.~\ref{sec:4} is devoted to concluding remarks.

\section{\label{sec:2}Generalized dispersion relation of the relativistic\\
current filamentation instability}

\subsection{\label{sec:2a}Basic equations and assumptions}

Begin with the nonlinear beam-Maxwell equations
that include both friction and pressure terms.
Assuming the ions to be at rest and to provide a
uniform charge-neutralizing background, we study
the relativistic dynamics of two uniform, counterstreaming electron currents
by employing the following set of equations in the dimensionless form,

\begin{widetext}
\begin{equation}
{\partial n_a \over {\partial t}} - \nabla \cdot {\bf j}_a = 0,
\label{eq:1}
\end{equation}

\begin{equation}
{\partial {\bf p}_a \over {\partial t}}
+ \left( {\bf v}_a \cdot \nabla {\bf p}_a \right)
= -\left( {\bf E} + {\bf v}_a \times {\bf B} \right)
- \nu_{ei} {\bf p}_a - \nu_{ee}
\left( {\bf p}_a - {\bf p}_{\bar{a}} \right)
- {\nabla P_a \over n_{0,a}},
\label{eq:2}
\end{equation}

\begin{equation}
\nabla \times {\bf E} = - {\partial {\bf B} \over {\partial t} },
\label{eq:3}
\end{equation}

\begin{equation}
\nabla \times {\bf B} = {\partial {\bf E} \over {\partial t} }
+ \sum_{a} {\bf j}_a,
\label{eq:4}
\end{equation}

\begin{equation}
\nabla \cdot {\bf E} = 1 - \sum_{a} n_a,
\label{eq:5}
\end{equation}
\end{widetext}

\noindent
where ${\bf v}_a = {\bf p}_a/\sqrt{1+p_a^2}$, ${\bf j}_a=-n_a v_a$,
and the subscript $a=1$, $2$ labels the two electron components
and $\bar{a}$ labels the countercomponent of $a$,
viz., $\bar{a} = 2$, $1$ for $a = 1$, $2$, respectively.
In particular,
$\nu_{ei}$ and $\nu_{ee}$ describe the electron-ion and electron-electron
momentum exchanges, and the other notations are standard.
For normalization, I have used the initially uniform density $n_0$,
the speed of light $c$, and the electron plasma frequency
$\omega_{pe} = \sqrt{4\pi n_0e^2/m_0}$.
Note that the Poisson Eq.~(\ref{eq:5}) is equivalent to a
combination of the continuity Eq.~(\ref{eq:1})
and the Ampere-Maxwell Eq.~(\ref{eq:4}).

According to a procedure similar to that developed by Califano {\it et al.}
\cite{califano97},
I investigate the behavior of small amplitude perturbations
by linearizing Eqs.~(\ref{eq:1})$-$(\ref{eq:4}).
I impose current neutrality of
$\sum_{a} n_{0,a}v_{0,a}=0$, where ${\bf v}_{0,a}=v_{0,a}\hat{{\bf x}}$
are the initial velocities in $x$ direction.
Under the current neutrality, there exists no magnetic field initially.
The CFI is studied in the $x$-$y$ plane.
In order to derive the dispersion relation,
all perturbed quantities are assumed to be in the form of
${\cal F}(y,t)={\cal F}{\rm exp}[i(k_y y - \omega t)]$.
As a result, a magnetic field ${\bf B}=(0,0,B_z)$ and
a corresponding electric field ${\bf E}=(E_x,E_y,0)$ are generated.
The purely transverse mode is investigated in detail throughout this paper,
i.e., $k_x=0$ and $k_y = k \neq 0$, whereas the longitudinal mode such as
two-stream instability with $k_x \neq 0$ and $k_y=0$ is
not taken into account at the moment.
The pressure of each electron components is connected with its density
by a polytropic relation,
which depends on the characteristic frequency $\omega$ and
wave number $k$ of the mode being considered.
As it is well known, in the case that the ratio of $\omega/k$ is
much larger than the electron thermal speed,
the adiabatic exponent of $\gamma_e = 3$
is adequate for the polytrope \cite{motz79,dendy90}.
Henceforth, I assumed $P_a/n_a^3 = {\rm const}$,
so that $\nabla P_a = 3 T_{0,a} \nabla n_a$.
Then, we get a closed form of the linearized Eqs.~(\ref{eq:1})$-$(\ref{eq:4}).
The background ions are supposed to be fixed on the short time scale
of $\sim \omega_{pe}^{-1}$.
Along these assumptions, in the nonrelativistic limit
Eqs.~(\ref{eq:1})$-$(\ref{eq:5}) involve the
resistive transverse wave modes and longitudinal wave mode with
thermal correction which were discussed in Ref.~\cite{motz79}.

\subsection{\label{sec:2b}The dispersion relation
including collisional effects and thermal corrections}

The extended dispersion relation of the relativistic CFI
including collisional and thermal effects is then found self-consistently.
In collisionless cases the dispersion relation that can be expressed
as functions of $\omega^2$ and $(k/\omega)^2$ does not include
the imaginary unit $i = \sqrt{-1}$ explicitly and
contains the purely real and purely imaginary solutions of $\omega(k)$
\cite{califano97}.
The purely real solutions consist of the pairs of
the positive and negative solutions,
corresponding to purely oscillatory and/or purely oscillatory wave modes,
while the purely imaginary solutions are concomitant with the
complex conjugate solutions,
to yield purely growing and purely decaying (damped) modes.

In the collisional case considered here,
the dispersion relation includes the imaginary unit explicitly.
Hence, the solutions of $\omega(k)$ may depart
from the real and imaginary axis in the complex plane.
In this sense, hereafter we refer to such solutions, i.e.,
complex eigenmodes with real and imaginary part, as {\it dephasing modes}.
Below, I explicitly write down the generalized dispersion relation
containing the dephasing modes.
After some manipulations outlined in the Appendix,
the dispersion relation can be obtained in the complex form of
${\rm Re}[\omega, (k/\omega)^2]+i{\rm Im}[\omega, (k/\omega)^2]=0$,
where

\begin{widetext}
\begin{subequations}
\label{eq:6}
$$
{\rm Re}[\omega, (k/\omega)^2] =
\left[
\left( 1 + \tau \Omega_k^{-2} \right)
- \left( \Omega_{11}^{-2} + \Omega_{11,T}^{-2} \Omega_k^{-2} \right)
\right]
$$
$$
\times
\left[
\left( 1 + \tau \Omega_k^{-2} \right) \left( 1 - \Omega_k^{-2} \right)
- \left( \Omega_{21}^{-2} + \Omega_{21,T}^{-2} \Omega_k^{-2} \right)
- \left( \Omega_{31}^{-2} + \Omega_{31,T}^{-2} \Omega_k^{-2} \right)
\Omega_k^{-2}
\right]
$$
$$
- \left( \Omega_{12}^{-2} + \Omega_{12,T}^{-2} \Omega_k^{-2} \right)
\left[
\left( \Omega_{22}^{-2} + \Omega_{22,T}^{-2} \Omega_k^{-2} \right)
+ \left( \Omega_{32}^{-2} + \Omega_{32,T}^{-2} \Omega_k^{-2} \right)
\Omega_k^{-2}
\right]
$$
\begin{equation}
- \left[
\left( \Omega_{41}^{-2} + \Omega_{41,T}^{-2} \Omega_k^{-2} \right)
\left( \Omega_{43}^{-2} + \Omega_{43,T}^{-2} \Omega_k^{-2} \right)
- \left( \Omega_{42}^{-2} + \Omega_{42,T}^{-2} \Omega_k^{-2} \right)
\left( \Omega_{44}^{-2} + \Omega_{44,T}^{-2} \Omega_k^{-2} \right)
\right]
\Omega_k^{-2},
\label{subeq:6a}
\end{equation}

$$
{\rm Im}[\omega, (k/\omega)^2] =
- \left( \Omega_{12}^{-2} + \Omega_{12,T}^{-2} \Omega_k^{-2} \right)
$$
$$
\times \left[
\left( 1 + \tau \Omega_k^{-2} \right) \left( 1 - \Omega_k^{-2} \right)
- \left( \Omega_{21}^{-2} + \Omega_{21,T}^{-2} \Omega_k^{-2} \right)
- \left( \Omega_{31}^{-2} + \Omega_{31,T}^{-2} \Omega_k^{-2} \right)
\Omega_k^{-2}
\right]
$$
$$
- \left[
\left( 1 + \tau \Omega_k^{-2} \right)
- \left( \Omega_{11}^{-2} + \Omega_{11,T}^{-2} \Omega_k^{-2} \right)
\right]
\left[
\left( \Omega_{22}^{-2} + \Omega_{22,T}^{-2} \Omega_k^{-2} \right)
+ \left( \Omega_{32}^{-2} + \Omega_{32,T}^{-2} \Omega_k^{-2} \right)
\Omega_k^{-2}
\right]
$$
\begin{equation}
- \left[
\left( \Omega_{41}^{-2} + \Omega_{41,T}^{-2} \Omega_k^{-2} \right)
\left( \Omega_{44}^{-2} + \Omega_{44,T}^{-2} \Omega_k^{-2} \right)
+ \left( \Omega_{42}^{-2} + \Omega_{42,T}^{-2} \Omega_k^{-2} \right)
\left( \Omega_{43}^{-2} + \Omega_{43,T}^{-2} \Omega_k^{-2} \right)
\right]
\Omega_k^{-2}.
\label{subeq:6b}
\end{equation}
\end{subequations}
\end{widetext}

\noindent
Here, $\tau$ is defined in Eq.~(\ref{subeq:9b}) later,
$\Omega_k^{-2} = (k/\omega)^2$, and furthermore,
$\Omega_{ij}^{-2}$ and $\Omega_{ij,T}^{-2}$ are defined by

\begin{widetext}
\begin{subequations}
\label{eq:7}
$$
\Omega_{11}^{-2} = \sum_{a}
{n_{0,a} \over {\Gamma_{0,a} \omega^2} } \phi_1,~~~
\Omega_{11,T}^{-2} = \sum_{a}
{n_{0,a} \over {\Gamma_{0,a} \omega^2} } \phi_{T,\bar{a}},
$$
\begin{equation}
\Omega_{12}^{-2} = \sum_{a}
{n_{0,a} \over {\Gamma_{0,a} \omega^2} } \psi_1,~~~
\Omega_{12,T}^{-2} = \sum_{a}
{n_{0,a} \over {\Gamma_{0,a} \omega^2} } \psi_{T,\bar{a}};
\label{subeq:7a}
\end{equation}
$$
\Omega_{21}^{-2} = \sum_{a}
{n_{0,a} \over {\Gamma_{0,a}^3 \omega^2} } \phi_2,~~~
\Omega_{21,T}^{-2} = \sum_{a}
{n_{0,a} \over {\Gamma_{0,a}^3 \omega^2} } \phi_{T},
$$
\begin{equation}
\Omega_{22}^{-2} = \sum_{a}
{n_{0,a} \over {\Gamma_{0,a}^3 \omega^2} } \psi_2,~~~
\Omega_{22,T}^{-2} = \sum_{a}
{n_{0,a} \over {\Gamma_{0,a}^3 \omega^2} } \psi_{T};
\label{subeq:7b}
\end{equation}
$$
\Omega_{31}^{-2} = \sum_{a}
{{n_{0,a}v_{0,a}^2} \over {\Gamma_{0,a} \omega^2} } \phi_{\bar{a},a},~~~
\Omega_{31,T}^{-2} = \sum_{a}
{{n_{0,a}v_{0,a}^2} \over {\Gamma_{0,a} \omega^2} } \phi_{T,\bar{a},a},
$$
\begin{equation}
\Omega_{32}^{-2} = \sum_{a}
{{n_{0,a}v_{0,a}^2} \over {\Gamma_{0,a} \omega^2} } \psi_{\bar{a},a},~~~
\Omega_{32,T}^{-2} = \sum_{a}
{{n_{0,a}v_{0,a}^2} \over {\Gamma_{0,a} \omega^2} } \psi_{T,\bar{a},a};
\label{subeq:7c}
\end{equation}
$$
\Omega_{41}^{-2} = \sum_{a}
{{n_{0,a}v_{0,a}} \over {\Gamma_{0,a} \omega^2} } \phi_1,~~~
\Omega_{41,T}^{-2} = \sum_{a}
{{n_{0,a}v_{0,a}} \over {\Gamma_{0,a} \omega^2} } \phi_{T,\bar{a}},
$$
$$
\Omega_{42}^{-2} = \sum_{a}
{{n_{0,a}v_{0,a}} \over {\Gamma_{0,a} \omega^2} } \psi_1,~~~
\Omega_{42,T}^{-2} = \sum_{a}
{{n_{0,a}v_{0,a}} \over {\Gamma_{0,a} \omega^2} } \psi_{T,\bar{a}},
$$
$$
\Omega_{43}^{-2} = \sum_{a}
{{n_{0,a}v_{0,a}} \over {\Gamma_{0,a} \omega^2} } \phi_{\bar{a},a},~~~
\Omega_{43,T}^{-2} = \sum_{a}
{{n_{0,a}v_{0,a}} \over {\Gamma_{0,a} \omega^2} } \phi_{T,\bar{a},a},
$$
\begin{equation}
\Omega_{44}^{-2} = \sum_{a}
{{n_{0,a}v_{0,a}} \over {\Gamma_{0,a} \omega^2} } \psi_{\bar{a},a},~~~
\Omega_{44,T}^{-2} = \sum_{a}
{{n_{0,a}v_{0,a}} \over {\Gamma_{0,a} \omega^2} } \psi_{T,\bar{a},a},
\label{subeq:7d}
\end{equation}
\end{subequations}
\end{widetext}

\noindent
where $\Gamma_{0,a} = (1 - v_{0,a}^2)^{-1/2}$ is the Lorentz factor, and
$\phi$ and $\psi$ stand for the dephasing factors which
can be expressed as

\begin{widetext}
$$
\phi_1 = { {\xi_1 + 2 \nu \nu_1} \over \phi},~~~
\phi_{T,\bar{a}}
= - { { \xi_1 \tilde{T}_{0,\bar{a}}
+ \left( 1 + \nu \nu_1 \right) T } \over \phi },
$$
$$
\psi_1 = { {2 \nu - \xi_1 \nu_1} \over \psi },~~~
\psi_{T,\bar{a}} = - { {\nu T_{\bar{a}} - \nu_1 T} \over \psi },
$$
$$
\phi_2 = {1 \over \zeta_2},~~~
\phi_T = - { {2 \xi_2 \tilde{T}} \over {\zeta_2 \phi} },
$$
$$
\psi_2 = - \tilde{\nu}_{ei} \phi_2,~~~
\psi_T = - \tilde{\nu}_{ei} \phi_T,
$$
$$
\phi_{\bar{a},a} =
{ {\xi_1 + 2\nu \left(\nu + \upsilon_{\bar{a}a} \tilde{\nu}_{ee} \right) }
\over \phi },~~~
\phi_{T,\bar{a},a} =
- { {\xi_1 \tilde{T}_{0,\bar{a}} + \left[
1 + \nu \left(\nu + \upsilon_{\bar{a}a} \tilde{\nu}_{ee} \right) \right]T }
\over \phi },
$$
$$
\psi_{\bar{a},a} =
{ { 2 \nu - \xi_1 \left( \nu + \upsilon_{\bar{a}a} \tilde{\nu}_{ee} \right) }
\over \psi },~~~
\psi_{T,\bar{a},a} =
- { {\nu T_{\bar{a}} - \left( \nu + \upsilon_{\bar{a}a} \tilde{\nu}_{ee}
\right)T } \over \psi },
$$
\begin{equation}
\phi = -\psi = \xi_1^2 + 4 \nu^2,
\label{eq:8}
\end{equation}
\end{widetext}

\noindent
and the abbreviations are

\begin{widetext}
\begin{subequations}
\label{eq:9}
$$
\nu = \tilde{\nu}_{ee} + \tilde{\nu}_{ei},
$$
$$
\nu_1 = \nu + \tilde{\nu}_{ee},
$$
$$
\zeta_1 = 1 + \tilde{\nu}_{ee}^2,~~~
\zeta_2 = 1 + \tilde{\nu}_{ei}^2,
$$
\begin{equation}
\xi_1 = \zeta_1 - \nu^2,~~~
\xi_2 = \zeta_1 + \nu^2;
\label{subeq:9a}
\end{equation}
\begin{equation}
\tau = \zeta_2 \phi_T,~~~
T = \sum_{a} \tilde{T}_{0,a},~~~
T_{\bar{a}} = 3 \tilde{T}_{0,\bar{a}} + \tilde{T}_{0,a};
\label{subeq:9b}
\end{equation}
\begin{equation}
\upsilon_{\bar{a} a} = {v_{0,\bar{a}} \over v_{0,a}},
\label{subeq:9c}
\end{equation}
\end{subequations}
\end{widetext}

\noindent
and $\tilde{\nu}_{ee} = \nu_{ee}/\omega$,
$\tilde{\nu}_{ei} = \nu_{ei}/\omega$, and
$\tilde{T}_{0,a} = 3 T_{0,a}/\Gamma_{0,a}$.
In collisionless limits, it follows that
in Eq.~(\ref{subeq:9a}) $\nu$, $\nu_1 \rightarrow 0$, and
$\zeta_1$, $\zeta_2$, $\xi_1$, $\xi_2 \rightarrow 1$;
for infinitesimal thermal spread, in Eq.~(\ref{subeq:9b})
$\tau$, $T$, $T_{\bar{a}} \rightarrow 0$; and
for symmetrically counterstreaming currents of
$v_{0,1} = -v_{0,2}$, in Eq.~(\ref{subeq:9c}) $\upsilon_{\bar{a}a} = -1$.
It is noted that the second order terms for thermal correction of the form of
$\sim (\tilde{T}_{0,a} \Omega_k^{-2})^2$
have been neglected, as explained in the Appendix.
The corresponding condition $\tilde{T}_{0,a} \Omega_k^{-2} \ll 1$
turns out to be consistent with the results obtained later,
as well as, with the aforementioned adiabatic condition of
$|k/\omega| v_{th,a} \ll 1$, where $v_{th,a} \approx \sqrt{T_{0,a}}$
is the electron thermal speed.
When assuming the normalized frequencies
$\tilde{\nu}_{ee}$ and $\tilde{\nu}_{ei}$ to be constants,
the dispersion Eq.~(\ref{eq:6}) can be expressed as
${\rm Re}(\omega^2, \Omega_k^{-2})+i{\rm Im}(\omega^2, \Omega_k^{-2})=0$,
which contains, in general, ten complex solutions of $\omega(k)$,
consisting of five pairs of positive and negative solutions.
In a special case, they may include purely real and/or
purely imaginary solutions.

\subsection{\label{sec:2c}Approximate dispersions for specific cases}

In this section, I investigate some specific cases
contained in the general result:
first for $\tilde{T}_{0,a} = 0$ with (1) only electron-electron collisions
and (2) only electron-ion collisions, and then for
(3) $\tilde{T}_{0,a} \neq 0$ and (4) $\tilde{T}_{0,a} = 0$,
without collisions.\\

\subsubsection{\label{sec:2c1}The case including electron-electron
collisional effects}

For the case of $\tilde{\nu}_{ee} \neq 0$,
$\tilde{\nu}_{ei} \rightarrow 0$, and $\tilde{T}_{0,a} \rightarrow 0$,
the dephasing factors of Eq.~(\ref{eq:8}) asymptotically lead to

\begin{widetext}
$$
\phi_1, \phi_2 \rightarrow 1;~~~
\phi_{\bar{a},a} \rightarrow
{ {1 + 2 \left(1 + \upsilon_{\bar{a}a} \right) \tilde{\nu}_{ee}^2 }
\over {1 + 4 \tilde{\nu}_{ee}^2 } };~~~
\phi_T, \phi_{T,\bar{a}}, \phi_{T,\bar{a},a} 
\rightarrow 0;
$$
\begin{equation}
\psi_1, \psi_2 \rightarrow 0;~~~
\psi_{\bar{a},a} \rightarrow
- { {
\left( 1 - \upsilon_{\bar{a}a} \right) \tilde{\nu}_{ee} }
\over {1 + 4 \tilde{\nu}_{ee}^2 } };~~~
\psi_T, \psi_{T,\bar{a}}, \psi_{T,\bar{a},a}
\rightarrow 0.
\label{eq:10}
\end{equation}
\end{widetext}

\noindent
Therefore, in Eq.~(\ref{eq:7}) we read
$\Omega_{12}^{-2}$, $\Omega_{22}^{-2}$, $\Omega_{42}^{-2}$,
$\Omega_{ij,T}^{-2} \rightarrow 0$, and Eq.~(\ref{eq:6}) then reduces to

\begin{widetext}
\begin{subequations}
\label{eq:11}
\begin{equation}
{\rm Re}(\omega^2, \Omega_k^{-2}) \approx
\left( 1 - \Omega_{11}^{-2} \right)
\left[
\left( 1 - \Omega_{21}^{-2} \right)
- \left( 1 + \Omega_{31}^{-2} \right) \Omega_k^{-2}
\right]
- \Omega_{41}^{-2} \Omega_{43}^{-2} \Omega_k^{-2},
\label{subeq:11a}
\end{equation}
\begin{equation}
{\rm Im}(\omega^2, \Omega_k^{-2}) \approx
- \left( 1 - \Omega_{11}^{-2} \right)
\Omega_{32}^{-2} \Omega_k^{-2}
- \Omega_{41}^{-2} \Omega_{44}^{-2} \Omega_k^{-2}.
\label{subeq:11b}
\end{equation}
\end{subequations}
\end{widetext}

\noindent
Instead of Eq.~(\ref{eq:7}), I introduce the definitions of

\begin{widetext}
\begin{subequations}
\label{eq:12}
\begin{equation}
\Omega_{1}^{-2} = \sum_{a}
{ n_{0,a} \over {\Gamma_{0,a} \omega^{2}} },~~
\Omega_{2}^{-2} = \sum_{a}
{ n_{0,a} \over {\Gamma_{0,a}^3 \omega^{2}} },~~
\Omega_{4}^{-2} = \sum_{a}
{ {n_{0,a}v_{0,a}} \over {\Gamma_{0,a} \omega^{2}} };
\label{subeq:12a}
\end{equation}
\begin{equation}
\Omega_{3}^{\prime -2} = \sum_{a}
{ {n_{0,a}v_{0,a}^2} \over {\Gamma_{0,a} \omega^{\prime 2}} },~~~
\Omega_{4}^{\prime -2} = \sum_{a}
{ {n_{0,a}v_{0,a}} \over {\Gamma_{0,a} \omega^{\prime 2}} },
\label{subeq:12b}
\end{equation}
\end{subequations}
\end{widetext}

\noindent
where $\omega^{\prime 2} = \omega^2/(\phi_{\bar{a},a} + i \psi_{\bar{a},a})$,
which may be rewritten as

\begin{widetext}
\begin{equation}
\omega^{\prime 2} =
{ {1 + 2 i \tilde{\nu}_{ee}} \over
{1 + i \left( 1 + \upsilon_{\bar{a}a} \right) \tilde{\nu}_{ee}} }
\omega^2.
\label{eq:13}
\end{equation}
\end{widetext}

\noindent
The definitions of $\Omega_{1}^{-2}$, $\Omega_{2}^{-2}$, and $\Omega_{4}^{-2}$
in Eq.~(\ref{subeq:12a}) are recalled later.
It is noted that for a trivial case of copropagating currents with
$v_{0,1} = v_{0,2}$, i.e., $\upsilon_{\bar{a}a} = 1$,
Eq.~(\ref{eq:13}) reduces to $\omega^{\prime 2} = \omega^2$,
indicating that indeed, dephasing effects vanish.
Making use of Eqs.~(\ref{eq:12}) and (\ref{eq:13}),
the approximate dispersion Eq.~(\ref{eq:11}) can be written in the form of

\begin{widetext}
\begin{equation}
\omega^2
\left( 1 - \Omega_1^{-2} \right)
\left( 1 - \Omega_2^{-2} \right)
- k^2
\left[
\left( 1 - \Omega_1^{-2} \right)
\left( 1 + \Omega_3^{\prime -2} \right) +
\Omega_4^{-2} \Omega_4^{\prime -2}
\right] \approx 0.
\label{eq:14}
\end{equation}
\end{widetext}

\noindent
Equation~(\ref{eq:14}) contains six solutions of $\omega$.
One should note that the factor of $(1 - \Omega_1^{-2})$,
regardless of the transformation of Eq.~(\ref{eq:13}),
involves a simple eigenmode of the relativistic plasma oscillation.
More details are discussed in Sec.~\ref{sec:3a}.

\subsubsection{\label{sec:2c2}The case including electron-ion
collisional effects}

For the case of $\tilde{\nu}_{ee} \rightarrow 0$,
$\tilde{\nu}_{ei} \neq 0$,
and $\tilde{T}_{0,a} \rightarrow 0$,
the dephasing factors of Eq.~(\ref{eq:8}) asymptotically lead to

\begin{widetext}
$$
\phi_1, \phi_{\bar{a},a} \rightarrow \phi_2
= { 1 \over \zeta_2 };~~~
\phi_T, \phi_{T,\bar{a}}, \phi_{T,\bar{a},a}
\rightarrow 0;
$$
\begin{equation}
\psi_1, \psi_{\bar{a},a} \rightarrow \psi_2
= - { \tilde{\nu}_{ei} \over \zeta_2 };~~~
\psi_T, \psi_{T,\bar{a}}, \psi_{T,\bar{a},a}
\rightarrow 0.\label{eq:15}
\end{equation}
\end{widetext}

\noindent
Therefore, in Eq.~(\ref{eq:7}) $\Omega_{ij,T}^{-2} \rightarrow 0$,
and Eq.~(\ref{eq:6}) then reduces to

\begin{widetext}
\begin{subequations}
\label{eq:16}
$$
{\rm Re}(\omega^2, \Omega_k^{-2}) \approx
\left( 1 - \Omega_{11}^{-2} \right)
\left[
\left( 1 - \Omega_{21}^{-2} \right)
- \left( 1 + \Omega_{31}^{-2} \right) \Omega_k^{-2}
\right]
$$
\begin{equation}
- \Omega_{12}^{-2}
\left( \Omega_{22}^{-2} + \Omega_{32}^{-2}\Omega_k^{-2} \right)
- \left( \Omega_{41}^{-2} \Omega_{43}^{-2}
- \Omega_{42}^{-2} \Omega_{44}^{-2} \right) \Omega_k^{-2},
\label{subeq:16a}
\end{equation}

$$
{\rm Im}(\omega^2, \Omega_k^{-2}) \approx
- \Omega_{12}^{-2}
\left[
\left( 1 - \Omega_{21}^{-2} \right)
- \left( 1 + \Omega_{31}^{-2} \right) \Omega_k^{-2}
\right]
$$
\begin{equation}
- \left( 1- \Omega_{11}^{-2} \right)
\left( \Omega_{22}^{-2} + \Omega_{32}^{-2} \Omega_k^{-2} \right)
- \left( \Omega_{41}^{-2} \Omega_{44}^{-2}
+ \Omega_{42}^{-2} \Omega_{43}^{-2} \right) \Omega_k^{-2}.
\label{subeq:16b}
\end{equation}
\end{subequations}
\end{widetext}

\noindent
Here, we note the relations of

\begin{widetext}
$$
\Omega_{12}^{-2} \approx - \tilde{\nu}_{ei} \Omega_{11}^{-2},~~~
\Omega_{22}^{-2} \approx - \tilde{\nu}_{ei} \Omega_{21}^{-2},~~~
\Omega_{32}^{-2} \approx - \tilde{\nu}_{ei} \Omega_{31}^{-2},~~~
$$
\begin{equation}
\Omega_{42}^{-2} \approx - \tilde{\nu}_{ei} \Omega_{41}^{-2},~~~
\Omega_{44}^{-2} \approx - \tilde{\nu}_{ei} \Omega_{43}^{-2},~~~
\Omega_{42}^{-2} \approx \Omega_{44}^{-2}.
\label{eq:17}
\end{equation}
\end{widetext}

\noindent
For heuristic ways, I properly choose the dephasing boost-frame defined as

\begin{widetext}
\begin{equation}
\omega^{\prime \prime 2} = (1 + i \tilde{\nu}_{ei}) \omega^2,
~~(\forall \omega^2);~~~~~
k^{\prime \prime 2} = (1 + i \tilde{\nu}_{ei}) k^2,
~~(\forall k^2).
\label{eq:18}
\end{equation}
\end{widetext}

\noindent
Note that in contrast with Eq.~(\ref{eq:13}),
the wave number is also transformed by the operator in Eq.~(\ref{eq:18}).
Taking account of the transformation, I give the definitions of

\begin{widetext}
$$
\Omega_{1}^{\prime \prime -2} = \sum_{a}
{ n_{0,a} \over {\Gamma_{0,a} \omega^{\prime \prime 2}} },~~
\Omega_{2}^{\prime \prime -2} = \sum_{a}
{ n_{0,a} \over {\Gamma_{0,a}^3 \omega^{\prime \prime 2}} },
$$
\begin{equation}
\Omega_{3}^{\prime \prime -2} = \sum_{a}
{ {n_{0,a}v_{0,a}^2} \over {\Gamma_{0,a} \omega^{\prime \prime 2}} },~~
\Omega_{4}^{\prime \prime -2} = \sum_{a}
{ {n_{0,a}v_{0,a}} \over {\Gamma_{0,a} \omega^{\prime \prime 2}} },
\label{eq:19}
\end{equation}
\end{widetext}

\noindent
instead of Eq.~(\ref{eq:7}).
By using Eqs.~(\ref{eq:18}) and (\ref{eq:19}),
the approximate dispersion Eq.~(\ref{eq:16}) can be expressed as

\begin{widetext}
\begin{equation}
\omega^{\prime \prime 2}
\left( 1 - \Omega_1^{\prime \prime -2} \right)
\left( 1 - \Omega_2^{\prime \prime -2} \right)
- k^{\prime \prime 2}
\left[
\left( 1 - \Omega_1^{\prime \prime -2} \right)
\left( 1 + \Omega_3^{\prime \prime -2} \right) +
\Omega_4^{\prime \prime -4}
\right] \approx 0,
\label{eq:20}
\end{equation}
\end{widetext}

\noindent
where $\Omega_4^{\prime \prime -4} = (\Omega_{4}^{\prime \prime -2})^2$.
Equation~(\ref{eq:20}) contains six solutions.
It is noted that the form of Eq.~(\ref{eq:20}) is quite similar to that of
Eq.~(\ref{eq:14}), though there exist some differences in dephasing property
between the two.
For example, the eigenmode relevant to the plasma oscillation,
which is contained in the factor $(1 - \Omega_1^{\prime \prime -2})$,
is now undergoing the transformation of Eq.~(\ref{eq:18}).
More on these is given later in Sec.~\ref{sec:3b} and \ref{sec:3e}.

\subsubsection{\label{sec:2c3}The case including thermal corrections}

For the case of $\tilde{\nu}_{ee}$, $\tilde{\nu}_{ei} \rightarrow 0$,
and $\tilde{T}_{0,a} \neq 0$,
the dephasing factors of Eq.~(\ref{eq:8}) asymptotically lead to

\begin{widetext}
$$
\phi_1, \phi_2, \phi_{\bar{a},a} \rightarrow 1;~~~
\phi_T \rightarrow \tau = -2T;~~~
\phi_{T,\bar{a}}, \phi_{T,\bar{a},a}
\rightarrow -\left( T + \tilde{T}_{0,\bar{a}} \right);
$$
\begin{equation}
\psi_1, \psi_2, \psi_{\bar{a},a},
\psi_T, \psi_{T,\bar{a}}, \psi_{T,\bar{a},a}
\rightarrow 0.\label{eq:21}
\end{equation}
\end{widetext}

\noindent
Therefore, in Eq.~(\ref{eq:7}), $\Omega_{i2}^{-2}$, $\Omega_{i2,T}^{-2}$,
$\Omega_{44}^{-2}$, $\Omega_{44,T}^{-2} \rightarrow 0$,
and Eq.~(\ref{eq:6}) then reduces to

\begin{widetext}
\begin{subequations}
\label{eq:22}
$$
{\rm Re}(\omega^2, \Omega_k^{-2}) \approx
\left[
\left( 1 + \tau \Omega_k^{-2} \right)
- \left( \Omega_{11}^{-2} + \Omega_{11,T}^{-2} \Omega_k^{-2} \right)
\right]
$$
$$
\times
\left[
\left( 1 + \tau \Omega_k^{-2} \right) \left( 1 - \Omega_k^{-2} \right)
- \left( \Omega_{21}^{-2} + \Omega_{21,T}^{-2} \Omega_k^{-2} \right)
- \left( \Omega_{31}^{-2} + \Omega_{31,T}^{-2} \Omega_k^{-2} \right)
\Omega_k^{-2}
\right]
$$
\begin{equation}
-
\left( \Omega_{41}^{-2} + \Omega_{41,T}^{-2} \Omega_k^{-2} \right)
\left( \Omega_{43}^{-2} + \Omega_{43,T}^{-2} \Omega_k^{-2} \right)
\Omega_k^{-2},
\label{subeq:22a}
\end{equation}
\begin{equation}
{\rm Im}(\omega^2, \Omega_k^{-2}) \approx 0,
\label{subeq:22b}
\end{equation}
\end{subequations}
\end{widetext}

\noindent
and we have the relations of

\begin{widetext}
\begin{equation}
\Omega_{21,T}^{-2} \approx \tau \Omega_{21}^{-2},~~~
\Omega_{41}^{-2} \approx \Omega_{43}^{-2},~~~
\Omega_{41,T}^{-2} \approx \Omega_{43,T}^{-2}.
\label{eq:23}
\end{equation}
\end{widetext}

\noindent
In this case, dephasing effects disappear, and the dispersion relation
yields purely real and purely imaginary solutions.
Concerning Eq.~(\ref{eq:23}), I give the definitions of

\begin{widetext}
\begin{subequations}
\label{eq:24}
\begin{equation}
\Omega_{1}^{-2} = \sum_{a}
{ n_{0,a} \over {\Gamma_{0,a} \omega^{2}} },~~~
\Omega_{1,T}^{-2} = \sum_{a}
{ n_{0,a} \over {\Gamma_{0,a} \omega^2} } \phi_{T,\bar{a}};
\label{subeq:24a}
\end{equation}
\begin{equation}
\Omega_{2}^{-2} = \sum_{a}
{ n_{0,a} \over {\Gamma_{0,a}^3 \omega^{2}} },~~~
\Omega_{2,T}^{-2} = \tau \Omega_{2}^{-2};
\label{subeq:24b}
\end{equation}
\begin{equation}
\Omega_{3}^{-2} = \sum_{a}
{ {n_{0,a}v_{0,a}^2} \over {\Gamma_{0,a} \omega^{2}} },~~~
\Omega_{3,T}^{-2} = \sum_{a}
{{n_{0,a}v_{0,a}^2} \over {\Gamma_{0,a} \omega^2} } \phi_{T,\bar{a},a};
\label{subeq:24c}
\end{equation}
\begin{equation}
\Omega_{4}^{-2} = \sum_{a}
{ {n_{0,a}v_{0,a}} \over {\Gamma_{0,a} \omega^{2}} },~~~
\Omega_{4,T}^{-2} = \sum_{a}
{{n_{0,a}v_{0,a}} \over {\Gamma_{0,a} \omega^2} } \phi_{T,\bar{a}},
\label{subeq:24d}
\end{equation}
\end{subequations}
\end{widetext}

\noindent
instead of Eq.~(\ref{eq:7}).
Note that the definitions of $\Omega_{1}^{-2}$, $\Omega_{2}^{-2}$, and
$\Omega_{4}^{-2}$ have already appeared in Eq.~(\ref{subeq:12a}).
Using the definitions of Eq.~(\ref{eq:24}),
the approximate dispersion Eq.~(\ref{eq:22}) can be written as

\begin{widetext}
$$
\left[
\left( 1 + \tau \Omega_k^{-2} \right)
- \left( \Omega_{1}^{-2} + \Omega_{1,T}^{-2} \Omega_k^{-2} \right)
\right]
$$
$$
\times \left[
\left( 1 + \tau \Omega_k^{-2} \right)
\left( 1 - \Omega_{2}^{-2} - \Omega_k^{-2} \right)
- \left( \Omega_{3}^{-2} + \Omega_{3,T}^{-2} \Omega_k^{-2} \right)
\Omega_k^{-2}
\right]
$$
\begin{equation}
-
\left( \Omega_{4}^{-2} + \Omega_{4,T}^{-2} \Omega_k^{-2} \right)^2
\Omega_k^{-2} \approx 0.
\label{eq:25}
\end{equation}
\end{widetext}

\noindent
In general, Eq.~(\ref{eq:25}) contains ten solutions.
At first glance, the form of Eq.~(\ref{eq:25}) seems to be
different from that of Eqs.~(\ref{eq:14}) and (\ref{eq:20}).
Indeed, there are some differences in dispersive property, for instance,
the plasma oscillation mode contained in
the factor of $[( 1 + \tau \Omega_k^{-2} )
- ( \Omega_{1}^{-2} + \Omega_{1,T}^{-2} \Omega_k^{-2} )]$
involves the thermal dispersion.
In the special case without thermal corrections, however,
Eq.~(\ref{eq:25}) recovers the same form with Eq.~(\ref{eq:20}),
as shown below.

\subsubsection{\label{sec:2c4}The collisionless case without thermal
corrections}

For the case of $\tilde{\nu}_{ee}$, $\tilde{\nu}_{ei} \rightarrow 0$,
and $\tilde{T}_{0,a} \rightarrow 0$,
the dephasing factors of Eq.~(\ref{eq:8}) asymptotically lead to

\begin{widetext}
$$
\phi_1, \phi_2, \phi_{\bar{a},a} \rightarrow 1;~~~
\phi_T, \phi_{T,\bar{a}}, \phi_{T,\bar{a},a} \rightarrow 0;
$$
\begin{equation}
\psi_1, \psi_2, \psi_{\bar{a},a},
\psi_T, \psi_{T,\bar{a}}, \psi_{T,\bar{a},a}
\rightarrow 0.\label{eq:26}
\end{equation}
\end{widetext}

\noindent
Therefore, in Eq.~(\ref{eq:7}),
$\Omega_{i2}^{-2}$, $\Omega_{44}^{-2}$,
$\Omega_{ij,T}^{-2} \rightarrow 0$ and
$\Omega_{41}^{-2} \approx \Omega_{43}^{-2}$, and then
Eq.~(\ref{eq:6}) reduces to

\begin{widetext}
\begin{equation}
\omega^{2}
\left( 1 - \Omega_1^{-2} \right)
\left( 1 - \Omega_2^{-2} \right)
- k^{2}
\left[
\left( 1 - \Omega_1^{-2} \right)
\left( 1 + \Omega_3^{-2} \right) +
\Omega_4^{-4}
\right] \approx 0,
\label{eq:27}
\end{equation}
\end{widetext}

\noindent
where $\Omega_{i}^{-2}$ have been defined in Eq.~(\ref{eq:24}),
and $\Omega_4^{-4} = (\Omega_4^{-2})^2$.
It is found that Eq.~(\ref{eq:27}) maintains the form similar to
Eqs.~(\ref{eq:14}) and (\ref{eq:20}).
This dispersion Eq.~(\ref{eq:27}) exactly coincides with that
obtained by Califano {\it et al.} \cite{califano97,califano98a}.

\section{\label{sec:3}Eigenmode properties and growth rates of the\\
relativistic current filamentation instability in a\\
collisional plasma}

In the following, we find the solutions contained in the approximate dispersion
Eqs.~(\ref{eq:14}), (\ref{eq:20}), (\ref{eq:25}), and (\ref{eq:27}).
The complex eigenmodes are explicitly written down,
and surveyed for wide parameter ranges of counterstreaming
relativistic currents.

\subsection{\label{sec:3a}Eigenmodes including electron-electron
collisional effects}

At first, I seek the solutions of Eq.~(\ref{eq:14}) in terms of $\omega$.
Let us consider the symmetrical configuration of counterstreaming currents
such as $n_{0,1} = n_{0,2} = 0.5$ \cite{tatarakis03},
having $v_{0,1} = -v_{0,2}$.
The choice of the parameters may be instructive for making a direct comparison
between the present results and the previous ones \cite{califano97}.
Equation~(\ref{eq:14}) can be then cast to

\begin{widetext}
\begin{equation}
\left( \omega^2 - {1 \over \Gamma_0} \right)
\left[
\omega^{\prime 2}
\left( \omega^2 - {1 \over \Gamma_0^3} \right) -
\left( \omega^{\prime 2} + {v_0^2 \over \Gamma_0} \right) k^2
\right] = 0,
\label{eq:28}
\end{equation}
\end{widetext}

\noindent
where $v_0 = |v_{0,a}|$ and $\Gamma_0 = (1 - v_0^2)^{-1/2}$, and
the transformation Eq.~(\ref{eq:13}) reduces to

\begin{widetext}
\begin{equation}
\omega^{\prime 2} = \left( 1 + 2 i \tilde{\nu}_{ee} \right) \omega^{2}.
\label{eq:29}
\end{equation}
\end{widetext}

\noindent
The first factor of (left-hand side) (lhs) of Eq.~(\ref{eq:28})
yields a simple electrostatic mode, corresponding to the relativistic
plasma oscillation: $\omega = \pm \omega_r$ and $\omega_i = 0$, where

\begin{widetext}
\begin{equation}
\omega_r(\Gamma_0) = \Gamma_0^{-1/2}.
\label{eq:30}
\end{equation}
\end{widetext}

\noindent
Below, we refer to this eigenmode as oscillatory mode ({\it o} mode).
It turns out that momentum exchange between electrons and electrons
does not disturb the plasma oscillation.

The second factor of lhs of Eq.~(\ref{eq:28})
contains four solutions.
The values of $\omega^{\prime 2}$ are connected with $\omega^{2}$
through the complex operator in Eq.~(\ref{eq:29}).
Therefore, in contrast to collisionless cases,
$\omega^{2}$ values are of complex, to give

\begin{widetext}
\begin{equation}
\omega^{2}(k^2, \tilde{\nu}_{ee}, \Gamma_0, \mu) = {1 \over 2 }
\left[
k^2 + {1 \over \Gamma_0^3} +
{\rm sgn}(\mu) \sqrt{ A(k^2, \tilde{\nu}_{ee}, \Gamma_0)}~{\rm exp}
\left[ i \vartheta(k^2, \tilde{\nu}_{ee}, \Gamma_0) \right]
\right],
\label{eq:31}
\end{equation}
\end{widetext}

\noindent
where ${\rm sgn}(\mu=\mu_\pm) = \pm 1$, and 

\begin{widetext}
$$
A(k^2, \tilde{\nu}_{ee}, \Gamma_0) =
\sqrt{ B^2(k^2, \tilde{\nu}_{ee}^2, \Gamma_0) +
C^2(k^2, \tilde{\nu}_{ee}, \Gamma_0) },
$$

$$
B(k^2, \tilde{\nu}_{ee}^2, \Gamma_0) =
k^4 + { {2 \left(2 \Gamma_0^2 + 4 \tilde{\nu}_{ee}^2 - 1 \right)} \over
\left( 1 + 4 \tilde{\nu}_{ee}^2 \right) \Gamma_0^3 } k^2 +
{ 1 \over \Gamma_0^6 },
$$

$$
C(k^2, \tilde{\nu}_{ee}, \Gamma_0) =
- { {8 \tilde{\nu}_{ee} \left( \Gamma_0^2 - 1 \right)} \over 
\left( 1 + 4 \tilde{\nu}_{ee}^2 \right) \Gamma_0^3 } k^2,
$$

\begin{equation}
\vartheta(k^2, \tilde{\nu}_{ee}, \Gamma_0) = {1 \over 2} \tan^{-1}
\left[
{
{ C(k^2, \tilde{\nu}_{ee}, \Gamma_0)
} \over {
B(k^2, \tilde{\nu}_{ee}^2, \Gamma_0) }
}
\right].
\label{eq:32}
\end{equation}
\end{widetext}

\noindent
Note the relation of $A+B > 0$ and $A-B \geq 0$.

Moreover, I rewrite Eq.~(\ref{eq:31})
in the form of the polar coordinate:
$\omega = \pm |\omega_0| {\rm exp}(i \theta_0)$,
where $\omega_0$ and $\theta_0$ are purely real numbers.
For the signs ($\pm$) each, we have two solutions with
$\mu=\mu_+$ and $\mu_-$, which are referred to as positive mode
({\it p} mode) and negative mode ({\it n} mode), respectively.
The squared magnitude and the polar angle of the complex eigenmodes
are given by

\begin{widetext}
\begin{subequations}
\label{eq:33}
\begin{equation}
\omega_0^2(k^2, \tilde{\nu}_{ee}, \Gamma_0, \mu) = { 1 \over 2 }
\sqrt{ \left[
k^2 + {1 \over \Gamma_0^3} + {\rm sgn}(\mu)
\alpha(k^2, \tilde{\nu}_{ee}, \Gamma_0)
\right]^2 +
\beta^2(k^2, \tilde{\nu}_{ee}, \Gamma_0) },
\label{subeq:33a}
\end{equation}

\begin{equation}
\theta_0(k^2, \tilde{\nu}_{ee}, \Gamma_0, \mu)
= { 1 \over 2 } {\rm tan}^{-1}
\left[ {
{ - {\rm sgn}(\mu)
\beta(k^2, \tilde{\nu}_{ee}, \Gamma_0) }
\over
{ k^2 + { 1 \over \Gamma_0^3 } + {\rm sgn}(\mu)
\alpha(k^2, \tilde{\nu}_{ee}, \Gamma_0) }
}
\right],
\label{subeq:33b}
\end{equation}
\end{subequations}
\end{widetext}

\noindent
respectively, where

\begin{widetext}
$$
\alpha(k^2, \tilde{\nu}_{ee}, \Gamma_0) =
\sqrt{
{ A(k^2, \tilde{\nu}_{ee}, \Gamma_0) +
B(k^2, \tilde{\nu}_{ee}^2, \Gamma_0) } \over 2 } > 0,
$$
\begin{equation}
\beta(k^2, \tilde{\nu}_{ee}, \Gamma_0) =
\sqrt{
{ A(k^2, \tilde{\nu}_{ee}, \Gamma_0) -
B(k^2, \tilde{\nu}_{ee}^2, \Gamma_0) } \over 2 } \geq 0.
\label{eq:34}
\end{equation}
\end{widetext}

\noindent
One should note that Eq.~(\ref{subeq:33b}) restricts its parameter range to
$-\pi/4 < \theta_0 \leq 0$,
because the argument of right-hand side (rhs) of Eq.~(\ref{subeq:33b}) is
negative definite, i.e., the ratio is 
$-/+$ for the {\it p} mode and $+/-$ for the {\it n} mode.
In order to reconstruct the polar angles
consistent with the argument, I appropriately define
$\omega_{0p} = |\omega_0(\mu=\mu_+)|$ and
$\theta_{0p} = \theta_0(\mu=\mu_+)$ for the {\it p} mode, and
$\omega_{0n} = |\omega_0(\mu=\mu_-)|$ and
$\theta_{0n} = \theta_0(\mu=\mu_-) + \pi/2$ for the {\it n} mode.
Note the allowable parameter ranges of
$-\pi/4 < \theta_{0p} \leq 0$ and
$\pi/4 < \theta_{0n} \leq \pi/2$.
For the {\it p} mode with
$-\sqrt{2}/2 < {\rm sin} \theta_{0p} \leq 0$ and
$\sqrt{2}/2 < {\rm cos} \theta_{0p} \leq 1$,
the complex eigenmodes of $\omega = \omega_{0p} {\rm exp}(i \theta_{0p})$
and $\omega = -\omega_{0p} {\rm exp}(i \theta_{0p})
= \omega_{0p} {\rm exp}[i (\theta_{0p}+\pi)]$
reflect decaying and growing waves, respectively.
Thus, the phase of the decaying and growing waves
propagates toward $y$ and $-y$ direction, respectively.
On the other hand, for the {\it n} mode with
$\sqrt{2}/2 < {\rm sin} \theta_{0n} \leq 1$ and
$0 \leq {\rm cos} \theta_{0n} < \sqrt{2}/2$,
the complex eigenmodes of $\omega = \omega_{0n} {\rm exp}(i \theta_{0n})$
and $\omega = -\omega_{0n} {\rm exp}(i \theta_{0n})
= \omega_{0n} {\rm exp}[i (\theta_{0n}-\pi)]$
reflect growing and decaying waves, respectively.
In contrast to the {\it p} mode, the phase of the growing and decaying waves
propagates toward $y$ and $-y$ direction, respectively.

In Fig.~\ref{fig:1} for the typical current speed of $v_0 = 0.9$
($\Gamma_0 = 2.29$),
I show the polar coordinate plots of the complex eigenmodes $\omega$
for given wave numbers $k$, varying electron-electron
collision parameter $\tilde{\nu}_{ee}$.
The trajectories of $\omega$ can be compared to those of
the arrowhead of vectors.
In the collisionless limit of $\tilde{\nu}_{ee} \rightarrow 0$,
we read $A \approx B$ and $C \rightarrow 0$ in Eq.~(\ref{eq:32}), and
$\beta \rightarrow 0$ in Eq.~(\ref{eq:34}),
so that $\theta_0 \rightarrow 0$ in Eq.~(\ref{subeq:33b}).
Hence, the vectors of the {\it p} mode direct the real axis,
reflecting the purely oscillatory wave mode
($\theta_{0p} = 0$; $\theta_{0p}+\pi = \pi$), whereas
the vectors of the {\it n} mode direct the imaginary axis,
reflecting the purely growing ($\theta_{0n} = \pi/2$) and
purely decaying ($\theta_{0n}-\pi = -\pi/2$) mode.
The unstable mode just represents the electromagnetic Weibel instability
in a collisionless plasma, which is discussed in Sec.~\ref{sec:3d} later.

As shown in Fig.~\ref{fig:1}(a) for $k = 3 \times 10^{-3}$,
for the small but finite value of $\tilde{\nu}_{ee}$
the vectors of the {\it n} mode depart clockwise from the imaginary axis.
As $\tilde{\nu}_{ee}$ increases, the real components of the vectors
increase, while the imaginary components decrease.
In this aspect, such a growing wave mode is
considered to be the dephasing Weibel mode with reduced growth rate.
In the strong collisional regime,
the polar angles of the vectors of the growing and decaying mode
approach $\theta_{0n} = \pi/4$ and
$\theta_{0n} - \pi = -3\pi/4$, respectively,
and the vectors shrink, reducing both their real and imaginary components.
The vectors of the {\it p} mode do not largely depart from the real axis
for the small value of $k$.

As seen in Fig.~\ref{fig:1}(b),
for the moderate value of $k = 3 \times 10^{-1}$,
the trajectories of the {\it n} mode are similar to those
for the smaller $k$ value.
As for the vectors of the {\it p} mode,
I now find the clockwise deviation from the real axis.
This represents the decaying ($\theta_{0p} < 0$) and
growing ($\theta_{0p}+\pi < \pi$) electromagnetic mode.
In the moderate collisional regime, the magnitude of
the deviation angles tends to be large,
though the maximum value is found to be small,
compared with that for the {\it n} mode.
In the strong collisional regime,
the vectors of the {\it p} mode are likely to return to the real axis,
reducing their magnitude.
For comparison, the fixed vectors of the {\it o} mode are also plotted
in the Fig.~\ref{fig:1}~(a,b).
In these cases, the magnitude of the vectors is larger than that of
the {\it p} and {\it n} mode.

For the {\it p} and {\it n} mode,
the angular frequency of the oscillations can be defined by
$\omega_r (\mu=\mu_+) = \omega_{0p} {\rm cos} \theta_{0p} > 0$ and
$\omega_r (\mu=\mu_-)= \omega_{0n} {\rm cos} \theta_{0n} \geq 0$,
respectively; and the linear growth rate can be defined by
$\gamma (\mu=\mu_+) = -\omega_{0p} {\rm sin} \theta_{0p} \geq 0$ and
$\gamma (\mu=\mu_-) = \omega_{0n} {\rm sin} \theta_{0n} > 0$, respectively.
They are summarized as follows:

\begin{widetext}
\begin{subequations}
\label{eq:35}
\begin{equation}
\omega_r(k^2, \tilde{\nu}_{ee}, \Gamma_0, \mu) =
{1 \over 2} \sqrt{ 2 \omega_0^2(k^2, \tilde{\nu}_{ee}, \Gamma_0)
+ \left[
k^2 + {1 \over \Gamma_0^3} + {\rm sgn}(\mu)
\alpha(k^2, \tilde{\nu}_{ee}, \Gamma_0)
\right] },
\label{subeq:35a}
\end{equation}
\begin{equation}
\gamma(k^2, \tilde{\nu}_{ee}, \Gamma_0, \mu) =
{1 \over 2} \sqrt{ 2 \omega_0^2(k^2, \tilde{\nu}_{ee}, \Gamma_0)
- \left[
k^2 + {1 \over \Gamma_0^3} + {\rm sgn}(\mu)
\alpha(k^2, \tilde{\nu}_{ee}, \Gamma_0)
\right] }.
\label{subeq:35b}
\end{equation}
\end{subequations}
\end{widetext}

\noindent
Here, note the relation of $\omega_r(\mu_+) > \gamma(\mu_+)$ and
$\omega_r(\mu_-) < \gamma(\mu_-)$.

\begin{figure*}
\resizebox{170mm}{!}{\includegraphics{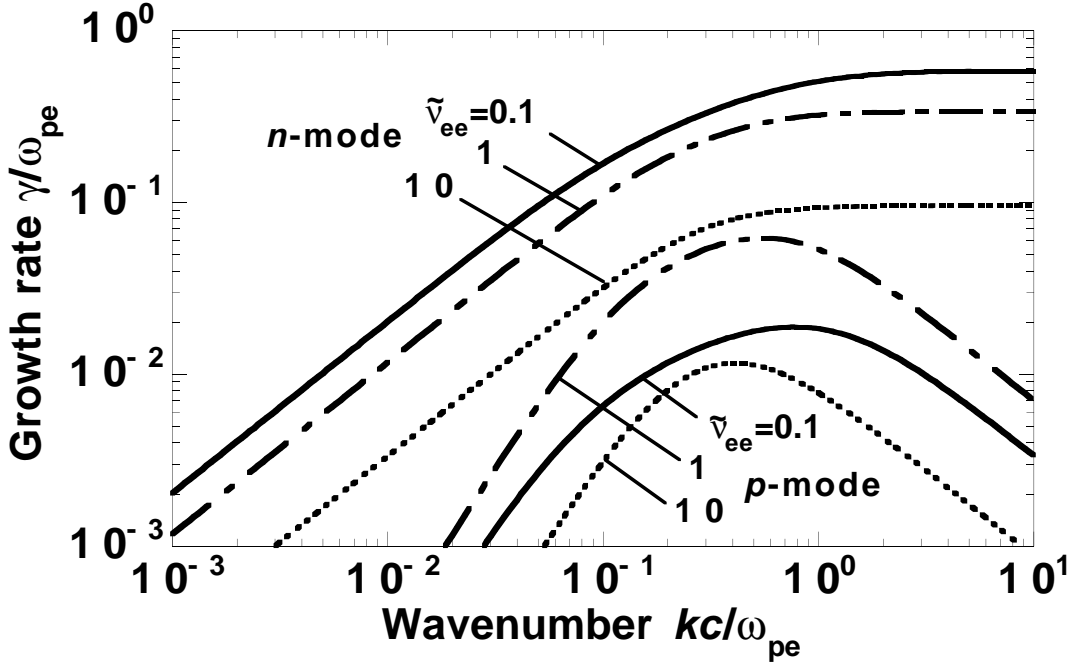}}
\caption{\label{fig:2}
The linear growth rate $\gamma$ of Eq.~(\ref{subeq:35b}) including the
effects of electron-electron collision, as a function of the wave number $k$,
for $\tilde{\nu}_{ee} = 0.1$ (solid curves),
$1$ (dot-dashed curves), and $10$ (dotted curves).
The {\it p} and {\it n} mode refer to the unstable modes for
$\mu=\mu_+$ and $\mu_-$ in Eq.~(\ref{subeq:35b}), respectively.
Here, I have chosen the parameter of $v_0 = 0.9$,
setting $\tilde{\nu}_{ei} = 0$ and $\tilde{T}_{0,a} = 0$.
}
\end{figure*}

In Fig.~\ref{fig:2} for $v_0 = 0.9$,
I show the growth rate $\gamma$ of Eq.~(\ref{subeq:35b})
as a function of the wave number $k$
for given collision parameters $\tilde{\nu}_{ee} = 0.1$, $1$, and $10$.
For the {\it n} mode, it is found that the collisional effects
lower the growth rate for all $k$,
maintaining the dependence of $\gamma \propto k$ for small value of $k$,
as well as, the saturation property for large value of $k$.
Such asymptotic properties of the growth rate can be also seen in the
purely growing Weibel mode in
a collisionless plasma \cite{califano97, califano98a}.
In addition, the collisional effects create the {\it p} mode
due to the dephasing mechanism mentioned above.
The growth rate has the dependence of $\gamma \propto k^2$ and
$\gamma \propto k^{-1}$ in the small- and large-$k$ region,
respectively, taking a peak around the moderate value of $k$.
Such a peak tends to be prominent for $\tilde{\nu}_{ee} \sim {\rm O}(1)$,
and then decreases as $\tilde{\nu}_{ee}$ further increases,
as consistent with Fig.~\ref{fig:1}(b).
As a result, the growth rate cannot exceed that of the {\it n} mode
corresponding to the dephasing Weibel mode with reduced growth rate.
Within the present framework,
it seems that both modes do not definitely cut off the growth of
short wavelength perturbations.

In Fig.~\ref{fig:3}, for $v_0 = 0.9$ and $\tilde{\nu}_{ee} = 1$,
I show the growth rate $\gamma$ as a function of the angular frequency
$\omega_r$ of Eq.~(\ref{subeq:35a}), varying the wave number $k$
as a parameter.
The growth rate of the {\it n} mode
turns out to be larger than that of the {\it p} mode for all $k$,
which is in contrast with the case including only
electron-ion collisional effects, as shown in Fig.~\ref{fig:8} later.
Moreover, I found that the angular frequencies of the
{\it p} and {\it n} mode are separated at
$\omega_r = \Gamma_0^{-3/2} \simeq 0.29$, and
the oscillation frequency of the {\it p} mode is
always higher than that of the {\it n} mode.
The separation frequency, where
$\partial \omega_r/ \partial k \rightarrow +0$, is lower than
the frequency of Eq.~(\ref{eq:30}) for the {\it o} mode,
i.e. the plasma cutoff frequency.
The relation of $\partial \omega_r/ \partial k > 0$
for both the {\it p} and {\it n} mode ensures that
the direction of the group velocities coincides with that
of the phase velocities of carrier wave.
For large value of $k$, the oscillation frequency of the {\it p} mode
goes far beyond the relativistic plasma frequency,
while the growth rate decreases.
These properties again appear in the case including electron-ion
collisional effects (see also Fig.~\ref{fig:8}).

\begin{figure*}
\resizebox{170mm}{!}{\includegraphics{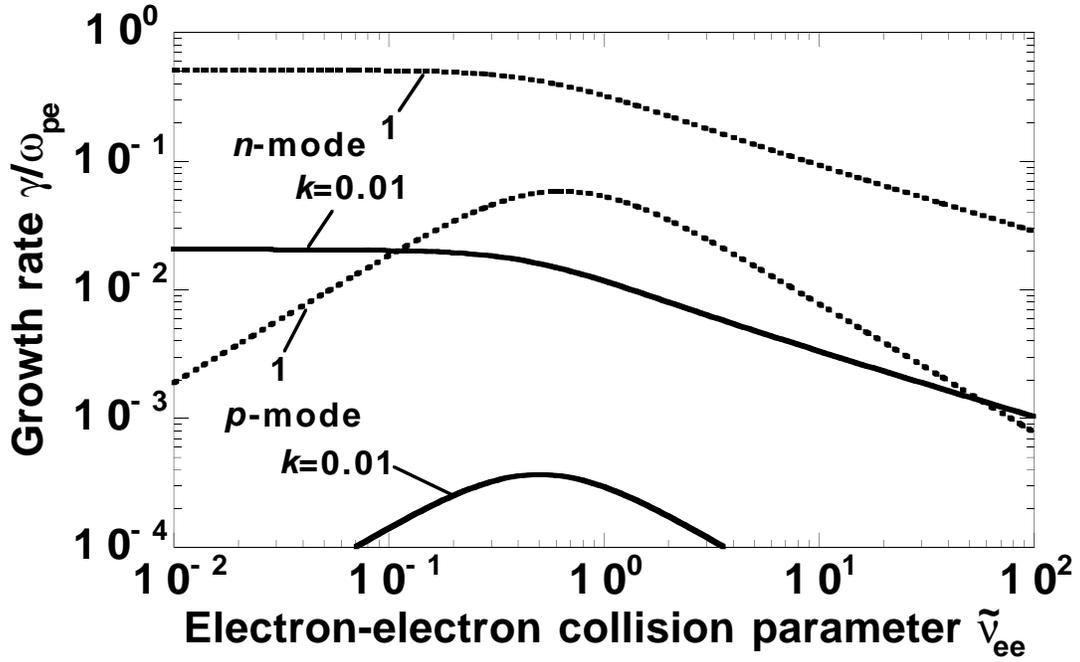}}
\caption{\label{fig:4}
The linear growth rate $\gamma$ for the {\it p} and {\it n} mode
given by Eq.~(\ref{subeq:35b}) as a function of $\tilde{\nu}_{ee}$
for $k = 0.01$ (solid curves) and $1$ (dotted curves).
Here, I have chosen the parameter of $v_0 = 0.9$,
setting $\tilde{\nu}_{ei} = 0$ and $\tilde{T}_{0,a} = 0$.
}
\end{figure*}

In Fig.~\ref{fig:4} for $v_0 = 0.9$,
I show the growth rate $\gamma$ as a function of
the collision parameter $\tilde{\nu}_{ee}$
for given values of $k = 0.01$ and $1$.
It is found that for moderate value of $\tilde{\nu}_{ee}$,
the growth rate of the {\it p} mode takes the peak value,
which tends to be well pronounced especially for $k \sim {\rm O}(1)$.
For example, for  $k = 1$ the growth rate takes the peak of
$\gamma \simeq 5.8 \times 10^{-2}$ at $\tilde{\nu}_{ee} \simeq 0.63$.
In the weak collisional regime, it has dependence of
$\gamma \propto \tilde{\nu}_{ee}$, but on the other hand
that of the {\it n} mode is almost constant, to give, e.g.,
$\gamma \simeq 0.51$ for $k = 1$.
In the strong collisional regime,
the growth rates of both the {\it p} and {\it n} mode decrease,
to exhibit the asymptotic behaviors of
$\gamma \propto \tilde{\nu}_{ee}^{-1}$ and
$\gamma \propto \tilde{\nu}_{ee}^{-0.5}$, respectively.

\begin{figure*}
\resizebox{170mm}{!}{\includegraphics{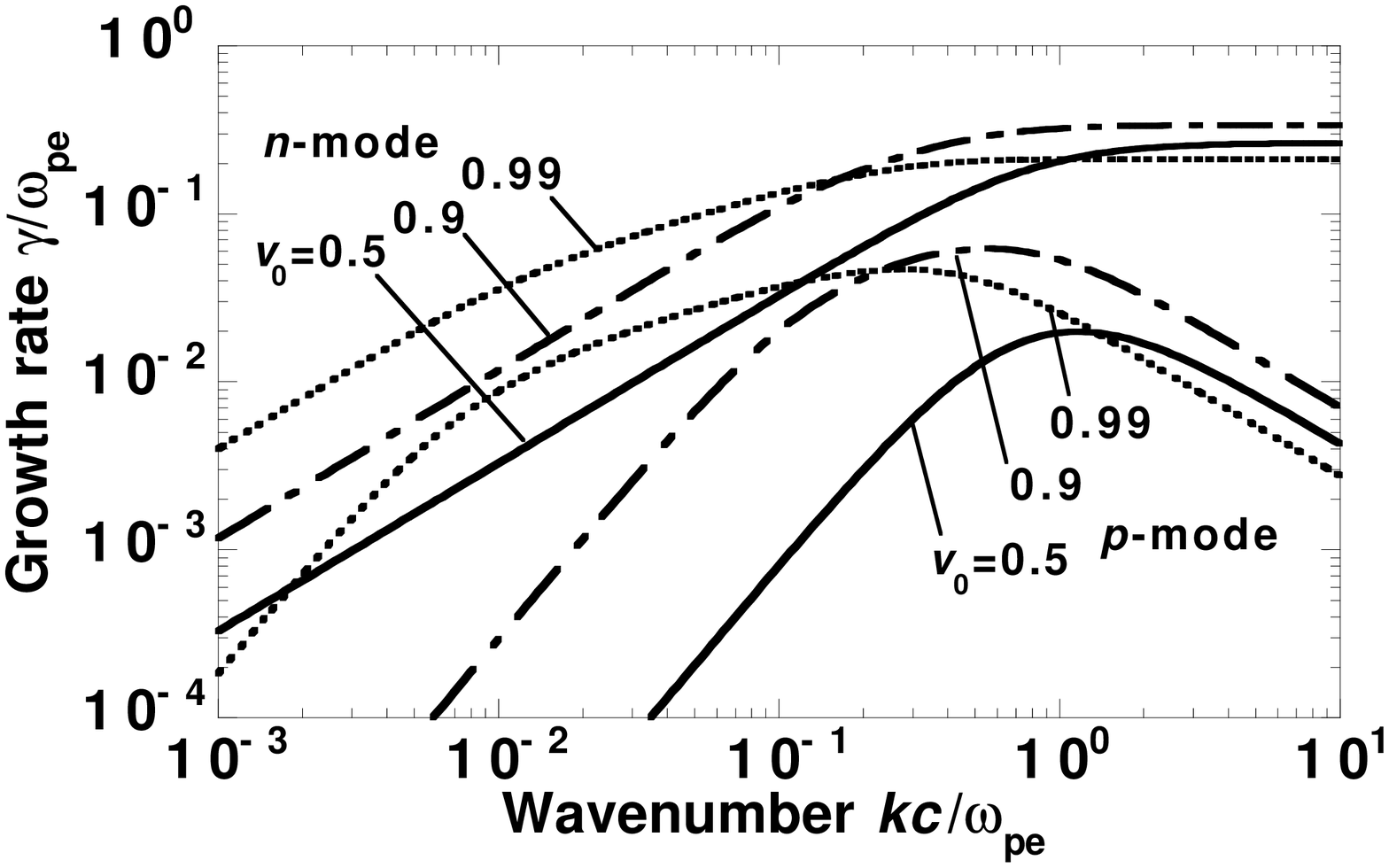}}
\caption{\label{fig:5}
The linear growth rate $\gamma$ for the {\it p} and {\it n} mode
given by Eq.~(\ref{subeq:35b}) as a function of the wave number $k$
for $v_0 = 0.5$ (solid curve), $0.9$ (dot-dashed curves),
and $0.99$ (dotted curve).
Here, I have chosen the parameter of $\tilde{\nu}_{ee} = 1$,
setting $\tilde{\nu}_{ei} = 0$ and $\tilde{T}_{0,a} = 0$.
Note that the dot-dashed curves for $v_0 = 0.9$ are the same as those
shown in Fig.~\ref{fig:2}.
}
\end{figure*}

In Fig.~\ref{fig:5} for $\tilde{\nu}_{ee} = 1$,
I show the growth rate $\gamma$ as a function of $k$
for given values of $v_0 = 0.5$, $0.9$, and $0.99$,
corresponding to $\Gamma_0 = 1.15$ ($79.0~{\rm keV}$),
$2.29$ ($611~{\rm keV}$), and $7.09$ ($3.11~{\rm MeV}$), respectively.
For small value of $k$,
both the {\it p} and {\it n} mode seem to increase their growth rates as
$\Gamma_0$ increases.
For large value of $k$, however,
such properties appear merely in weak to mild relativistic regime.
Namely, in strong relativistic regime,
the growth for large $k$ tends to be suppressed, for example,
as for the {\it n} mode, the saturation level of the growth rate
decreases as $\Gamma_0$ increases.
The peak of the growth rate of the {\it p} mode
is likely to shift to the smaller-$k$ region, reducing its value.
Although, the energy dependence of the {\it n} mode appears again
in the case including electron-ion collisional effects,
the {\it p} mode significantly changes its property, as shown below.

\subsection{\label{sec:3b}Eigenmodes including electron-ion
collisional effects}

I seek the solutions of Eq.~(\ref{eq:20}) in terms of $\omega$.
In the case of the symmetrically counterstreaming currents with
$n_{0,1} = n_{0,2} = 0.5$ and $v_{0,1} = -v_{0,2}$,
Eq.~(\ref{eq:20}) can be cast to

\begin{widetext}
\begin{equation}
\left( \omega^{\prime \prime 2} - {1 \over \Gamma_0} \right)
\left[
\omega^{\prime \prime 2}
\left( \omega^{\prime \prime 2} - {1 \over \Gamma_0^3} \right) -
\left( \omega^{\prime \prime 2} + {v_0^2 \over \Gamma_0} \right)
k^{\prime \prime 2}
\right] = 0,
\label{eq:36}
\end{equation}
\end{widetext}

\noindent
where $v_0 = |v_{0,a}|$ and $\Gamma_0 = (1 - v_0^2)^{-1/2}$.
Note that Eq.~(\ref{eq:36}) has the same form as Eq.~(\ref{eq:28})
for the previous case.
However, now all $\omega^2$ and $k^2$ values are being dephased by the
transformation Eq.~(\ref{eq:18}), to provide the eigenmodes significantly
different from those derived from Eq.~(\ref{eq:28}).
The first factor of lhs of Eq.~(\ref{eq:36}) contains
a modified electrostatic mode of the relativistic plasma oscillation.
The eigenmode may be written in the form of
$\omega = \pm (\omega_r - i \omega_i)$,
and the growth rate can be then defined by $\gamma = \omega_i$, that is,

\begin{widetext}
\begin{equation}
\omega_r(\tilde{\nu}_{ei}^2, \Gamma_0) =
{ 1 \over {\sqrt{\Gamma_0}}
\zeta_2^{3/4}(\tilde{\nu}_{ei}^2) },~~~
\gamma(\tilde{\nu}_{ei}, \Gamma_0) =
{ \tilde{\nu}_{ei} \over {\sqrt{\Gamma_0}
\zeta_2^{3/4}}(\tilde{\nu}_{ei}^2) }.
\label{eq:37}
\end{equation}
\end{widetext}

\noindent
It is found that the dissipative effects owing to the electron-ion collisions
create the growing and decaying oscillatory mode,
which {\it seemingly} carries its phase toward $-y$ and $y$ direction,
respectively.
As far as ignoring thermal and asymmetrical effects of
counterstreaming currents is concerned,
this mode is independent of wave number,
so that the group velocity is null.
Note that in the collisionless limit of $\tilde{\nu}_{ei} \rightarrow 0$,
Eq.~(\ref{eq:37}) reduces to Eq.~(\ref{eq:30})
which denotes the relativistic plasma oscillation.
In this sense, we refer to the eigenmode identified by Eq.~(\ref{eq:37}),
for convenience, as {\it o} mode.

The second factor of lhs of Eq.~(\ref{eq:36}) contains four
complex solutions.
Along the manner explained in Sec.~\ref{sec:3a}, I express
the solutions in the polar coordinate form of
$\omega = \pm |\omega_0| {\rm exp}(i \theta_0)$,
where $\omega_0$ and $\theta_0$ are purely real numbers.
For the signs ($\pm$) each, we get two solutions.
The squared magnitude and the polar angle of the complex solutions
are given by

\begin{widetext}
\begin{subequations}
\label{eq:38}
\begin{equation}
\omega_0^2(k^2, \tilde{\nu}_{ei}, \Gamma_0, \mu) =
{ 1 \over { 2 \zeta_2} }
\sqrt{
\left[
\zeta_2 k^2 + { 1 \over \Gamma_0^3 }
+ {\rm sgn}(\mu)
\left( \alpha + \tilde{\nu}_{ei} \beta \right)
\right]^2 +
\left[
{ \tilde{\nu}_{ei} \over \Gamma_0^3 }
+ {\rm sgn}(\mu)
\left( \tilde{\nu}_{ei} \alpha - \beta \right)
\right]^2  },
\label{subeq:38a}
\end{equation}

\begin{equation}
\theta_0(k^2, \tilde{\nu}_{ei}, \Gamma_0, \mu)
= { 1 \over 2 } {\rm tan}^{-1}
\left[ {
{
- { \tilde{\nu}_{ei} \over \Gamma_0^3 }
- {\rm sgn}(\mu)
\left( \tilde{\nu}_{ei} \alpha - \beta \right)
}
\over
{ \zeta_2 k^2 + { 1 \over \Gamma_0^3 }
+ {\rm sgn}(\mu)
\left( \alpha + \tilde{\nu}_{ei} \beta \right)
}
}
\right],
\label{subeq:38b}
\end{equation}
\end{subequations}
\end{widetext}

\noindent
respectively, where

\begin{widetext}
$$
\alpha(k^2, \tilde{\nu}_{ei}, \Gamma_0) =
\sqrt{
{ A(k^2, \tilde{\nu}_{ei}, \Gamma_0) +
B(k^2, \tilde{\nu}_{ei}^2, \Gamma_0) } \over 2 } > 0,
$$
\begin{equation}
\beta(k^2, \tilde{\nu}_{ei}, \Gamma_0) =
\sqrt{
{ A(k^2, \tilde{\nu}_{ei}, \Gamma_0) -
B(k^2, \tilde{\nu}_{ei}^2, \Gamma_0) } \over 2 } \geq 0,
\label{eq:39}
\end{equation}
\end{widetext}

\noindent
and,

\begin{widetext}
$$
A(k^2, \tilde{\nu}_{ei}, \Gamma_0) =
\sqrt{ B^2(k^2, \tilde{\nu}_{ei}^2, \Gamma_0) +
C^2(k^2, \tilde{\nu}_{ei}, \Gamma_0) },
$$

$$
B(k^2, \tilde{\nu}_{ei}^2, \Gamma_0) =
\left( 1 - \tilde{\nu}_{ei}^2 \right) k^4 +
{ { 2 \left( 2 \Gamma_0^2 - 1 \right) } \over { \Gamma_0^3 } } k^2 +
{ 1 \over \Gamma_0^6 },
$$

\begin{equation}
C(k^2, \tilde{\nu}_{ei}, \Gamma_0) = 2 \tilde{\nu}_{ei} k^2
\left( k^2 +
{ { 2 \Gamma_0^2 - 1 } \over { \Gamma_0^3 } }
\right).
\label{eq:40}
\end{equation}
\end{widetext}

\noindent
Note the relation of $A+B > 0$ and $A-B \geq 0$.
Below, I refer to the solutions with $\mu=\mu_+$ and $\mu_-$, as
{\it p} mode and {\it n} mode, respectively.
As is the case with Eq.~(\ref{subeq:33b}),
Eq.~(\ref{subeq:38b}) holds a parameter range of
$-\pi/4 < \theta_0 \leq 0$,
because the argument of rhs of Eq.~(\ref{subeq:38b}) is
negative definite,
i.e., the ratio is $-/+$ for the {\it p} mode and $+/-$ for the {\it n} mode.
Thus, recalling the definitions of
$\omega_{0p} = |\omega_0(\mu=\mu_+)|$ and
$\theta_{0p} = \theta_0(\mu=\mu_+)$ for the {\it p} mode, and
$\omega_{0n} = |\omega_0(\mu=\mu_-)|$ and
$\theta_{0n} = \theta_0(\mu=\mu_-) + \pi/2$ for the {\it n} mode,
the complex eigenmodes $\omega = \pm \omega_{0p} {\rm exp}(i \theta_{0p})$
denote the decaying ($+$) and growing ($-$) wave that
carry their phases toward $y$ and $-y$ direction, respectively;
and $\omega = \pm \omega_{0n} {\rm exp}(i \theta_{0n})$
denote the growing ($+$) and decaying ($-$) wave that
carry their phases toward $y$ and $-y$ direction, respectively.

In Fig.~\ref{fig:6} for the current speed of $v_0 = 0.9$,
I show the polar coordinate plots of the complex eigenmodes $\omega$
for given wave numbers $k$,
varying electron-ion collision parameter $\tilde{\nu}_{ei}$.
We compare the trajectories of $\omega$ to
those of the arrowhead of vectors.
In the collisionless limit of $\tilde{\nu}_{ei} \rightarrow 0$,
we read $A \approx B$ and $C \rightarrow 0$ in Eq.~(\ref{eq:40}), and
$\beta \rightarrow 0$ in Eq.~(\ref{eq:39}),
so that $\theta_0 \rightarrow 0$ in Eq.~(\ref{subeq:38b}).
Hence, the vectors of the {\it p} mode direct the real axis,
reflecting the purely oscillatory wave mode
($\theta_{0p} = 0$; $\theta_{0p}+\pi = \pi$), while
the vectors of the {\it n} mode direct the imaginary axis,
reflecting the purely growing ($\theta_{0n} = \pi/2$) and
purely decaying ($\theta_{0n}-\pi = -\pi/2$) mode.
The unstable mode represents the electromagnetic Weibel instability
in a collisionless plasma.

As shown in Fig.~\ref{fig:6}(a) for $k = 3 \times 10^{-3}$
for small but finite value of $\tilde{\nu}_{ei}$
the vectors of the {\it p} mode depart clockwise from the real axis.
As $\tilde{\nu}_{ei}$ increases, the real components of the vectors
decrease, while the imaginary components increase.
The magnitude of the deviation angles is larger than that
in the case including only electron-electron collisional effects
[compare Fig.~\ref{fig:1}(a)].
In the strong collisional regime, the vectors shrink,
reducing both their real and imaginary components.
This is the decaying and growing electromagnetic mode,
which possesses the allowed range of polar angle of
$-\pi/4 < \theta_{0p} \leq 0$ and
$3\pi/4 < \theta_{0p} + \pi \leq \pi$, respectively.
For the parameter range of $\tilde{\nu}_{ei}$ being considered,
the vectors of the {\it n} mode do not largely depart from
the imaginary axis, that is, the purely growing Weibel mode
is not so dephased.
The most remarkable property can be seen in the {\it o} mode.
For $\tilde{\nu}_{ei} \neq 0$, the vectors leave the real axis, and
as $\tilde{\nu}_{ei}$ increases
the magnitude of the deviation angles increases.
As consistent with Eq.~(\ref{eq:37}),
at $\tilde{\nu}_{ei} = 1$ the vectors of the growing
and decaying oscillatory mode have the angles of $3\pi/4$ and
$-\pi/4$, respectively;
and in the strong collisional limit,
asymptotically approach $\pi/2$ and $-\pi/2$, respectively.

In Fig.~\ref{fig:6}(b), for the moderate value of $k = 3 \times 10^{-1}$,
I now clearly find, for $\tilde{\nu}_{ei} \neq 0$,
the clockwise deviation of vector pairs of
{\it p}, {\it n}, and {\it o} modes.
This essentially means that all modes are in phase lag,
because of the frictional nature of collisions.
The vectors of the {\it p} and {\it n} mode
exhibit the behavior similar to that displayed in Fig.~\ref{fig:1}(b).
In contrast with the case for small $k$,
the polar angles of the vectors of the {\it p} mode cannot reach to
$3\pi/4$ and $-\pi/4$, and in the strong collisional regime
the vectors are likely to return to the real axis,
reducing their magnitude.
On the other hand, the vectors of the {\it n} mode
largely deviate from the imaginary axis until the angles reach to
$\pi/4$ and $-3\pi/4$.
In the weak collisional regime, the vectors increase the real components
and decrease the imaginary components, whereas
in the strong collisional regime, there is
decrease both in the real and imaginary components.
As a result,
the electron-ion collisional effects lower the growth rate
of the {\it n} mode, namely, the dephasing Weibel mode.
Note that the trajectories of the vectors of the {\it o} mode
are the same as those displayed in Fig.~\ref{fig:6}(a).
For the moderate value of $k$,
the magnitude of the vectors of the {\it n} mode is found to be
comparable to that of the {\it o} mode.

For the {\it p} and {\it n} mode,
the angular frequency of the oscillations can be defined by
$\omega_r (\mu=\mu_+) = \omega_{0p} {\rm cos} \theta_{0p} > 0$ and
$\omega_r (\mu=\mu_-)= \omega_{0n} {\rm cos} \theta_{0n} \geq 0$,
respectively; and the linear growth rate can be defined by
$\gamma (\mu=\mu_+) = -\omega_{0p} {\rm sin} \theta_{0p} \geq 0$ and
$\gamma (\mu=\mu_-) = \omega_{0n} {\rm sin} \theta_{0n} > 0$, respectively.
They are summarized as follows:

\begin{widetext}
\begin{subequations}
\label{eq:41}
\begin{equation}
\omega_r(k^2, \tilde{\nu}_{ei}, \Gamma_0, \mu) = { 1 \over 2 }
\sqrt{ 2 \omega_0^2(k^2, \tilde{\nu}_{ei}, \Gamma_0)
+ \left[ k^2 + {
{ { 1 \over \Gamma_0^3 }
+ {\rm sgn}(\mu)
\left[ \alpha(k^2, \tilde{\nu}_{ei}, \Gamma_0)
+ \tilde{\nu}_{ei}
\beta(k^2, \tilde{\nu}_{ei}, \Gamma_0)
\right]
} \over {\zeta_2(\tilde{\nu}_{ei}^2)} } \right] },
\label{subeq:41a}
\end{equation}

\begin{equation}
\gamma(k^2, \tilde{\nu}_{ei}, \Gamma_0, \mu) = { 1 \over 2 }
\sqrt{ 2 \omega_0^2(k^2, \tilde{\nu}_{ei}, \Gamma_0)
- \left[ k^2 + {
{ { 1 \over \Gamma_0^3 }
+ {\rm sgn}(\mu)
\left[ \alpha(k^2, \tilde{\nu}_{ei}, \Gamma_0)
+ \tilde{\nu}_{ei}
\beta(k^2, \tilde{\nu}_{ei}, \Gamma_0)
\right]
} \over {\zeta_2(\tilde{\nu}_{ei}^2)} } \right] }.
\label{subeq:41b}
\end{equation}
\end{subequations}
\end{widetext}

\noindent
Here, note the relation of $\omega_r(\mu_+) > \gamma(\mu_+)$ and
$\omega_r(\mu_-) < \gamma(\mu_-)$.

\begin{figure*}
\resizebox{170mm}{!}{\includegraphics{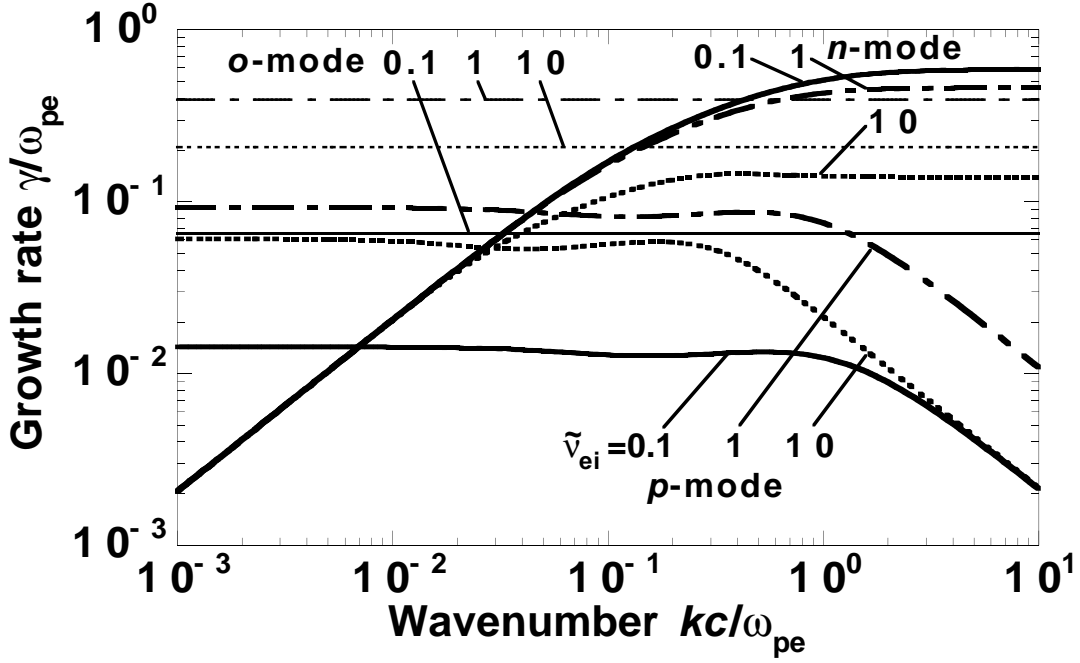}}
\caption{\label{fig:7}
The linear growth rate $\gamma$ of Eq.~(\ref{subeq:41b}) including the
effects of electron-ion collision, as a function of the wave number $k$,
for $\tilde{\nu}_{ei} = 0.1$ (solid curves),
$1$ (dot-dashed curves), and $10$ (dotted curves).
The {\it p} and {\it n} mode refer to the unstable modes for
$\mu=\mu_+$ and $\mu_-$ in Eq.~(\ref{subeq:41b}), respectively.
Hair lines for each $\tilde{\nu}_{ei}$ show the linear growth rate
$\gamma$ of Eq.~(\ref{eq:37}) for the {\it o} mode.
Here, I have chosen the parameters of $v_0 = 0.9$,
setting $\tilde{\nu}_{ee} = 0$ and $\tilde{T}_{0,a} = 0$.
}
\end{figure*}

In Fig.~\ref{fig:7} for $v_0 = 0.9$,
I show the growth rate $\gamma$ of Eqs.~(\ref{eq:37}) and (\ref{subeq:41b})
as a function of the wave number $k$
for given collision parameters $\tilde{\nu}_{ei} = 0.1$, $1$, and $10$.
For the {\it n} mode, the collisional effects lower the growth rate,
especially in the large-$k$ region where the growth rate tends to saturate.
In contrast to the case including only electron-electron collisions,
the growth rate for small $k$, which has the dependence of
$\gamma \propto k$, is found to be not so depressed
due to the electron-ion collisional effects (compare Fig.~\ref{fig:2}).
The {\it p} and {\it o} mode,
which appear owing to the collisional effects,
are prominent for the parameter of
$\tilde{\nu}_{ei} \sim {\rm O}(1)$, and sufficiently surpass the {\it n} mode
in growing long-wavelength (small-$k$) perturbations.
This feature is in major contrast to the case
including only electron-electron collisional effects.
For the larger collision parameter,
it is noteworthy that even in the large-$k$ region,
the growth rate of the {\it o} mode exceeds
that of the {\it n} mode, i.e., the suppressed Weibel mode.
This is one of the most important results in the present paper.
In the large-$k$ region, the growth rate of the {\it p} mode decreases,
showing the dependence of $\gamma \propto k^{-1}$, and
is far below that of the {\it n} and {\it o} mode.
In the small-$k$ region,
the growth rate of the {\it p} mode is almost independent of $k$,
and is always smaller than that of the {\it o} mode.
That is, $\gamma(\tilde{\nu}_{ei}, \Gamma_0, \mu=\mu_+) \approx
[(\sqrt{ \zeta_2 } - 1)/(2 \zeta_2 \Gamma_0^3)]^{1/2} <
\tilde{\nu}_{ei} / (\sqrt{\Gamma_0} \zeta_2^{3/4} )$
for $\tilde{\nu}_{ei} > 0$ and $\Gamma_0 > 1$.

In Fig.~\ref{fig:8}, for $v_0 = 0.9$ and $\tilde{\nu}_{ei} = 1$,
I show $\gamma$ of Eq.~(\ref{subeq:41b})
as a function of the angular frequency
$\omega_r$ of Eq.~(\ref{subeq:41a}), varying $k$ as a parameter.
For comparison, in the $\omega_r - \gamma$ plane,
I also plot a fixed point given by Eq.~(\ref{eq:37}) for the {\it o} mode.
It is found that for large value of $k$,
the growth rate of the {\it n} mode is larger than that of the {\it o} mode,
which is always larger than that of the {\it p} mode.
Note that, for the {\it p} and {\it n} mode,
$\partial \omega_r/ \partial k > 0$, and
the angular frequencies of both modes are clearly
separated at the frequency of $\omega_r \simeq 0.2$, where
$\partial \omega_r/ \partial k \rightarrow +0$.
As a result, the oscillation frequency of the {\it p} mode is
always higher than that of the {\it n} mode.
These features could be also seen in Fig.~\ref{fig:3} for the case
including electron-electron collisional effects.

\begin{figure*}
\resizebox{170mm}{!}{\includegraphics{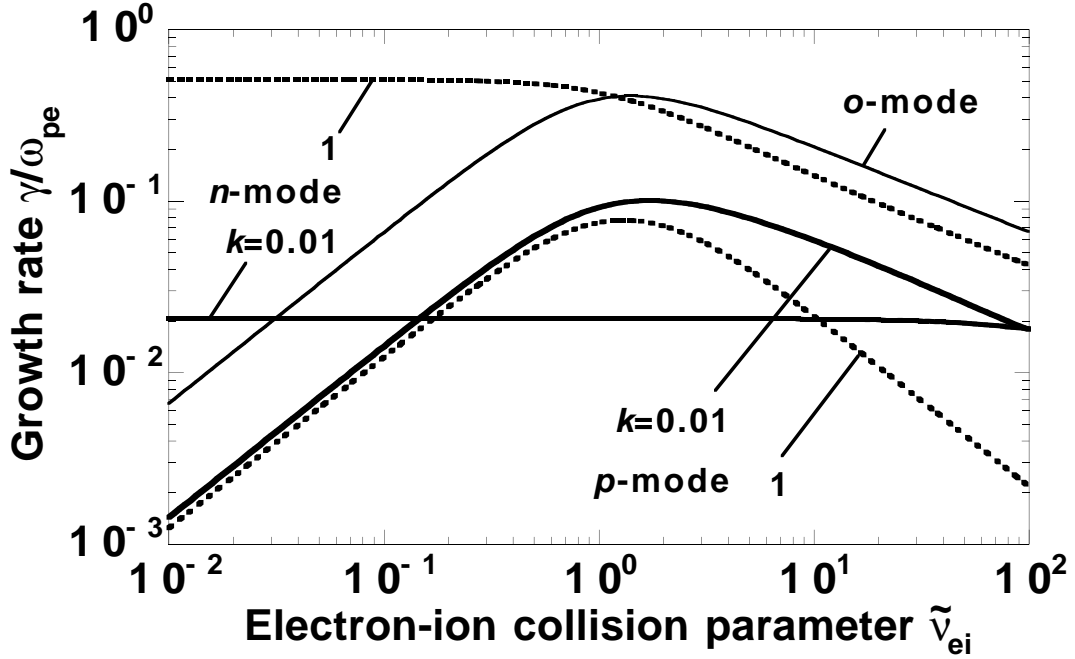}}
\caption{\label{fig:9}
The linear growth rate $\gamma$ for the {\it p} and {\it n} mode
given by Eq.~(\ref{subeq:41b}) as a function of $\tilde{\nu}_{ei}$
for $k = 0.01$ (solid curves) and $1$ (dotted curves).
Hair solid curve shows the linear growth rate $\gamma$ of
Eq.~(\ref{eq:37}) for the {\it o} mode, which is independent of $k$.
Here, I have chosen the parameter of $v_0 = 0.9$,
setting $\tilde{\nu}_{ee} = 0$ and $\tilde{T}_{0,a} = 0$.
}
\end{figure*}

In Fig.~\ref{fig:9}, for $v_0 = 0.9$,
I show the growth rate $\gamma$ of Eq.~(\ref{eq:37}), and (\ref{subeq:41b})
as a function of the collision parameter $\tilde{\nu}_{ei}$
for given values of $k = 0.01$ and $1$.
In the weak collisional regime,
the growth rates of the {\it o} and {\it p} mode are
both proportional to $\tilde{\nu}_{ei}$,
while the growth rate of the {\it n} mode is almost constant.
For smaller $k$, the growth rates of the {\it o} and {\it p} mode
can more readily exceed the growth rate of the {\it n} mode, and
for the moderate value of $\tilde{\nu}_{ei}$,
they take the peak values.
The growth rate of the {\it o} mode can, even for large $k$,
exceed that of the {\it n} mode, and takes the peak of
$\gamma \simeq 0.41$ at $\tilde{\nu}_{ei} \simeq 1.4$.
The growth rate decreases as $\tilde{\nu}_{ei}$ further increases,
showing the dependence of $\gamma \propto \tilde{\nu}_{ei}^{-0.5}$.
It is also noted that in the strong collisional regime,
the growth rates of the {\it p} and {\it n} mode
have the dependencies of $\gamma \propto \tilde{\nu}_{ei}^{-1}$
and $\gamma \propto \tilde{\nu}_{ei}^{-0.5}$, respectively.
As would be expected, anyhow,
all modes are suppressed in the strong collisional regime.

\begin{figure*}
\resizebox{170mm}{!}{\includegraphics{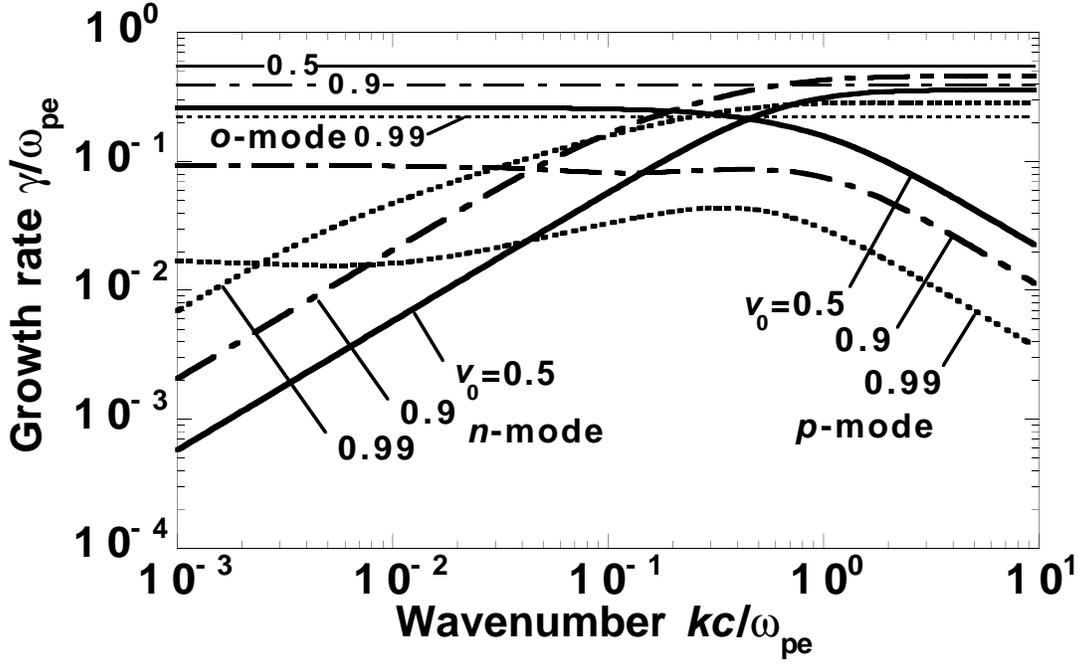}}
\caption{\label{fig:10}
The linear growth rate $\gamma$ for the {\it p} and {\it n} mode
given by Eq.~(\ref{subeq:41b}) as a function of the wave number $k$
for $v_0 = 0.5$ (solid curve), $0.9$ (dot-dashed curves),
and $0.99$ (dotted curve). Hair lines for each $v_0$ show
$\gamma$ of Eq.~(\ref{eq:37}) for the {\it o} mode.
Here, I have chosen the parameter of $\tilde{\nu}_{ei} = 1$,
setting $\tilde{\nu}_{ee} = 0$ and $\tilde{T}_{0,a} = 0$.
Note that the dot-dashed curves/line for $v_0 = 0.9$ are the same as those
shown in Fig.~\ref{fig:7}.
}
\end{figure*}

In Fig.~\ref{fig:10} for $\tilde{\nu}_{ei} = 1$,
I show the growth rate $\gamma$ of Eqs.~(\ref{eq:37}) and (\ref{subeq:41b})
as a function of $k$
for given values of $v_0 = 0.5$ ($\Gamma_0 = 1.15$),
$0.9$ ($\Gamma_0 = 2.29$), and $0.99$ ($\Gamma_0 = 7.09$).
It should be remarked that for the lower current speed,
the {\it o} mode becomes most dominant.
In contrast to the case including only electron-electron collisional effects,
the {\it o} and {\it p} mode monotonically reduce their growth rates
as $\Gamma_0$ increases,
and soon the {\it o} mode is overcome by the {\it n} mode
from large-$k$ region.
The energy dependence of the {\it n} mode is similar to that
shown in Fig.~\ref{fig:5},
namely, in weak to mild relativistic regime,
the saturation level of the growth rate
increases as $\Gamma_0$ increases, while in strong relativistic regime,
it tends to decrease.
In small-$k$ region, the growth rate seems to increase as
$\Gamma_0$ increases, since its curve, roughly proportional to $k$,
shifts to the smaller-$k$ region.
Such an apparent redshift can be also seen in the large-$k$ region
for the {\it p} mode.

\subsection{\label{sec:3c}Eigenmodes including thermal corrections}

In Eq.~(\ref{eq:25}), I seek the solutions in terms of $\omega$.
Regarding the symmetrically counterstreaming currents of
$n_{0,1} = n_{0,2} = 0.5$, $v_{0,1} = -v_{0,2}$, and
$\tilde{T}_{0,1} = \tilde{T}_{0,2}$, Eq.~(\ref{eq:25}) can be cast to

\begin{widetext}
$$
\left[
\omega^2 \left( \omega^2 - 4 \tilde{T}_0 k^2 \right) -
{1 \over \Gamma_0} \left( \omega^2 - 3 \tilde{T}_0 k^2 \right)
\right]
$$
\begin{equation}
\times \left[
\omega^2 \left( \omega^2 - 4 \tilde{T}_0 k^2 \right)
\left( \omega^2 - k^2 - {1 \over \Gamma_0^3} \right) -
{ v_0^2 \over \Gamma_0 } \left( \omega^2 - 3 \tilde{T}_0 k^2 \right) k^2
\right] = 0,
\label{eq:42}
\end{equation}
\end{widetext}

\noindent
where $v_0 = |v_{0,a}|$, $\Gamma_0 = (1-v_0^2)^{-1/2}$,
and $\tilde{T}_0 = \tilde{T}_{0,a}$.
The first factor of lhs of Eq.~(\ref{eq:42})
contains four purely real solutions,
but two of them are found to be inconsistent with
the assumption of $\tilde{T}_0 \Omega_k^{-2} \ll 1$ (not shown).
The other two solutions can be expressed as
$\omega = \pm \omega_r$, and
for $4 \Gamma_0 \tilde{T}_0 k^2\ll 1$, we obtain

\begin{widetext}
\begin{equation}
\omega_r(k^2, T_0, \Gamma_0) \approx
{ 1 \over {\sqrt{\Gamma_0}} } \left( 1 + { 3 \over 2} T_0 k^2 \right),
\label{eq:43}
\end{equation}
\end{widetext}

\noindent
where $T_0 = T_{0,a}$.
Note that Eq.~(\ref{eq:43}) is of the order of
$\omega_r \sim \Gamma_0^{-1/2}$,
and therefore, the assumption $\tilde{T}_0 \Omega_k^{-2} \ll 1$
requires $3 T_0 k^2 \ll 1$, which is consistent with the aforementioned
relation of $\sim \Gamma_0 \tilde{T}_0 k^2\ll 1$.
Equation~(\ref{eq:43}) just corresponds to the
relativistically extended dispersion of the Bohm-Gross wave
with nonzero group velocity
(e.g., see Ref.~\cite{motz79} for the nonrelativistic limit).
In contrast to Eqs.~(\ref{eq:30}) and (\ref{eq:37}),
we find the thermal dispersion terms characterized by $\sim T_0 k^2$
in Eq.~(\ref{eq:43}).
Physically, this reflects the Debye screening by electrons \cite{dendy90}.
In the limit of $T_0 k^2 \rightarrow 0$, Eq.~(\ref{eq:43}) reduces to
Eq.~(\ref{eq:30}) for oscillatory mode.

Moreover, the second factor of lhs of Eq.~(\ref{eq:42}) contains
six solutions.
Numerical calculation indicates that
these consist of four purely real solutions and a purely imaginary solution
concomitant with its complex conjugate: $\omega = \pm i \omega_i$.
For the purely growing mode that is of interest here,
we can define the linear growth rate as $\gamma = \omega_i$.
In Fig.~\ref{fig:11} (solid curve),
I show $\gamma$ as a function of the wave number $k$
for $v_0 = 0.9$ and $\tilde{T}_0 = 0.1$, as an example.
It is found that the thermal effects simply lower the growth rate
of the purely growing Weibel instability, at least,
in the range of $k \sim {\rm O}(1)$.
Note that in contrast to collisional cases, the thermal effects
do not take part in dephasing the purely oscillatory and purely growing mode,
but merely give rise to the mode-dispersion which can reduce the growth rate.

\subsection{\label{sec:3d}Eigenmodes of collisionless case without
thermal corrections}

In the collisionless limits without thermal corrections,
the dephasing operators in Eqs.~(\ref{eq:18}) and (\ref{eq:29})
asymptotically approach unity, vanishing their imaginary parts.
Then, Eqs.~(\ref{eq:28}), (\ref{eq:36}), and (\ref{eq:42})
degenerate into a unique equation,

\begin{widetext}
\begin{equation}
\left( \omega^{2} - {1 \over \Gamma_0} \right)
\left[
\omega^{2}
\left( \omega^{2} - {1 \over \Gamma_0^3} \right) -
\left( \omega^{2} + {v_0^2 \over \Gamma_0} \right) k^{2}
\right] = 0,
\label{eq:44}
\end{equation}
\end{widetext}

\noindent
where $v_0 = |v_{0,a}|$ and $\Gamma_0 = (1 - v_0^2)^{-1/2}$.
As expected,
Eq.~(\ref{eq:44}) has the same form with Eqs.~(\ref{eq:28}) and (\ref{eq:36}).
The first factor of lhs of Eq.~(\ref{eq:44}) yields
the plasma oscillation mode given by Eq.~(\ref{eq:30}).

The second factor of lhs of Eq.~(\ref{eq:44}) contains four solutions.
In contrast to collisional cases,
the values of $\omega^2$ are obtained as purely real numbers.
Concerning the polar form $\omega = \pm |\omega_0| {\rm exp}(i \theta_0)$,
Eqs.~(\ref{eq:33}) and (\ref{eq:38}) reduce to

\begin{widetext}
\begin{subequations}
\label{eq:45}

\begin{equation}
\omega_0^2(k^2, \Gamma_0, \mu) =
{ 1 \over 2 }
\left| k^2 + { 1 \over \Gamma_0^3 } + {\rm sgn}(\mu)
\sqrt{ A(k^2, \Gamma_0) } \right|,
\label{subeq:45a}
\end{equation}

\begin{equation}
\theta_0 = 0,
\label{subeq:45b}
\end{equation}
\end{subequations}
\end{widetext}

\noindent
where,

\begin{widetext}
\begin{equation}
A(k^2, \Gamma_0) = k^4 +
{ { 2 \left( 2 \Gamma_0^2 - 1 \right) } \over { \Gamma_0^3 } } k^2 +
{ 1 \over \Gamma_0^6 }.
\label{eq:46}
\end{equation}
\end{widetext}

\noindent
Recalling the definitions of
$\omega_{0p} = |\omega_0(\mu=\mu_+)|$ and $\theta_{0p} = \theta_0$
for the {\it p} mode, and $\omega_{0n} = |\omega_0(\mu=\mu_-)|$ and
$\theta_{0n} = \theta_0 + \pi/2$ for the {\it n} mode,
Eq.~(\ref{subeq:45b}) leads to
$\theta_{0p} = 0$ and $\theta_{0n} = \pi/2$.
It is, therefore, found that for the {\it p} mode,
the real solutions of
$\omega = \pm \omega_{0p} {\rm exp}(i \theta_{0p}) = \pm \omega_{0p}$
describe the purely oscillatory wave mode, while
for the {\it n} mode,
the imaginary solutions of $\omega = \pm \omega_{0n} {\rm exp}(i \theta_{0n})
= \pm i \omega_{0n}$
describe the purely growing ($+$) and purely decaying ($-$) mode.
The phase properties are shown in Figs.~\ref{fig:1} and \ref{fig:6}.

Now, the angular frequency of the oscillation and
the linear growth rate can be defined by $\omega_r = \omega_{0p} > 0$ and
$\gamma = \omega_{0n} > 0$, respectively.
Taking the inequality $\sqrt{A} > k^2 + (1/\Gamma_0^3)$
into consideration,
the angular frequency and the growth rate can be expressed as

\begin{figure*}
\resizebox{170mm}{!}{\includegraphics{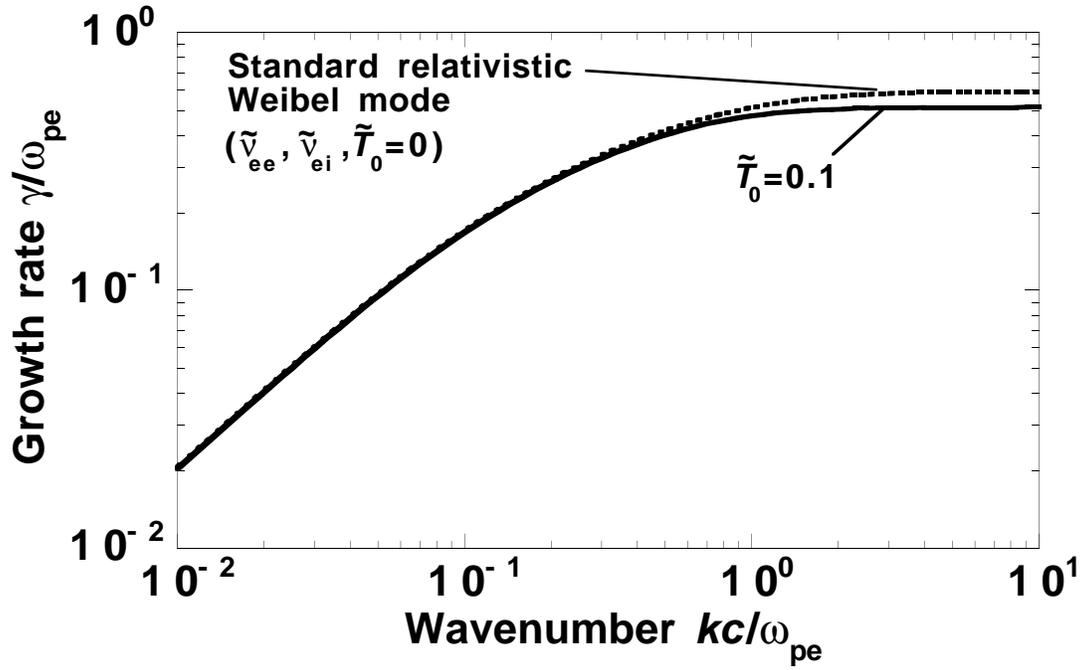}}
\caption{\label{fig:11}
The linear growth rate $\gamma$ as a function of the wave number $k$
for $\tilde{T}_0 = 0.1$ (solid curve) and $\tilde{T}_0 = 0$
[Eq.~(\ref{subeq:47b}): dotted curve] \cite{califano97}.
Here, I have chosen the parameters of $v_0 = 0.9$, $\tilde{\nu}_{ee} = 0$,
and $\tilde{\nu}_{ei} = 0$.
}
\end{figure*}

\begin{widetext}
\begin{subequations}
\label{eq:47}
\begin{equation}
\omega_r(k^2, \Gamma_0) =
{1 \over \sqrt{2} }
\left[ \sqrt{ A(k^2, \Gamma_0) } + \left(
k^2 + { 1 \over \Gamma_0^3 } \right) \right]^{1/2},
\label{subeq:47a}
\end{equation}

\begin{equation}
\gamma(k^2, \Gamma_0) =
{1 \over \sqrt{2} }
\left[ \sqrt{ A(k^2, \Gamma_0) } - \left(
k^2 + { 1 \over \Gamma_0^3 } \right) \right]^{1/2},
\label{subeq:47b}
\end{equation}
\end{subequations}
\end{widetext}

\noindent
respectively. Note the relation of $\omega_r > \gamma$.
In Fig.~\ref{fig:11} (dotted curve),
I show the linear growth rate $\gamma$ of Eq.~(\ref{subeq:47b})
for $v_0 = 0.9$, as a function of the wave number $k$.
This growth rate is just of the relativistically extended electromagnetic
Weibel instability in a collisionless plasma \cite{califano98b}.
It is found that for $\Gamma_0^5 k^2 \ll 1$ and $\Gamma_0^3 k^2 \gg 1$,
Eq.~(\ref{subeq:47b}) simply exhibits the asymptotic property
of $\gamma \approx \sqrt{\Gamma_0^2 - 1} k$ and
$\gamma \approx \sqrt{(\Gamma_0^2 - 1)/\Gamma_0^3}$, respectively.
It might be instructive to compare Fig.~\ref{fig:11} with
Figs.~\ref{fig:2} and \ref{fig:7} for the case including collisional effects.
It is confirmed that this mode is in a special case of the modes
presented in Sec.~\ref{sec:3a}$-$\ref{sec:3c}.

\subsection{\label{sec:3e}Eigenmodes including asymmetric effects of
counterstreaming currents}

\begin{figure*}
\resizebox{170mm}{!}{\includegraphics{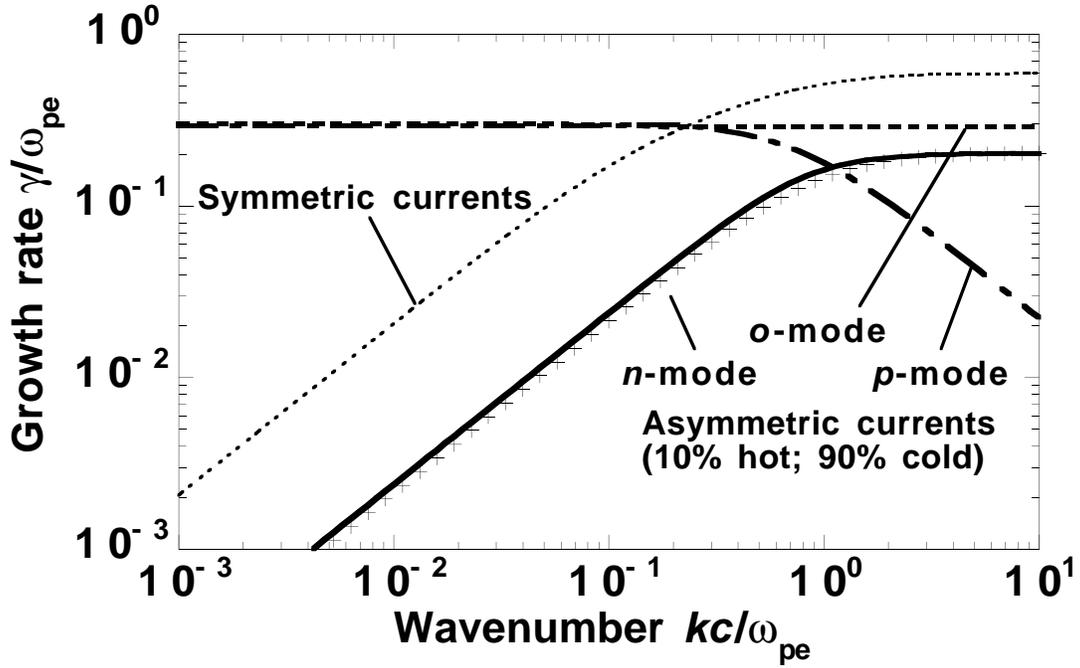}}
\caption{\label{fig:12}
The linear growth rate $\gamma$ as a function of the wave number $k$
for asymmetrically counterstreaming currents of
$v_{0,1} = 0.9$ and $v_{0,2} = -0.1$ with $\tilde{\nu}_{ei,2} = 0$
(crosses) and $\tilde{\nu}_{ei,2} = 1$.
For the latter case ($\tilde{\nu}_{ei,2} = 1$), solid, dashed, and dot-dashed
curves show $\gamma$ of the {\it n}, {\it o}, and {\it p} modes, respectively.
Here, I have set $\tilde{\nu}_{ei,1} = 0$ for both the cases.
For comparison, I also plot $\gamma$ for a symmetrical case of $v_0 = 0.9$
without collisions (hair dotted curve), which is the same as the
dotted curve shown in Fig.~\ref{fig:11}.
}
\end{figure*}

For another comparison, we are concerned with an asymmetrical configuration of
counterstreaming currents.
First, in Eq.~(\ref{eq:27}) for the collisionless case,
I change the parameters to an asymmetrical,
though still current-neutral initial beam configuration
with $n_{0,1} = 0.1$ ($v_{0,1} = 0.9$) and $n_{0,2} = 0.9$ ($v_{0,2} = -0.1$).
Note that the corresponding beam-to-plasma density ration, $n_b/n_p \sim 10$,
is of relevance to electron transport in the context of ignitor physics.
Here the electron beam density could be estimated as
$n_b \sim n_c \approx 10^{21}~{\rm cm^{-3}}$ and the plasma density as
$n_p \approx 10 n_c$ \cite{honda00c}.
It is around the low-density foot of steeply rising density profile
of laser-ablative plasma \cite{honda03a},
where the filamentation dynamics may be most prominent.
As shown in Fig.~\ref{fig:12} (crosses),
it is found that for such parameters
the growth rate reduces by a factor $10$ in the small-$k$ region
and by factor $\sqrt{10}$ in the saturation region.
The strong reduction is, more or less, favorable for
energetic electrons to propagate through the ablative corona
surrounding a highly compressed ignitor plasma.

In the denser region, electron-ion collisions might play an
important role in attenuating {\it slow} return currents,
owing to the relation of
$( \nu_{bi} / \omega_{pb} )( \nu_{ei} / \omega_{pe} )^{-1} \simeq
( n_b/n_e )^{5/2} \Gamma_b^{-3/2} \ll 1$,
where $\nu_{bi}$ and $\omega_{pb}$ denote the beam electron-ion collision
frequency and beam electron plasma frequency, respectively.
At this juncture, in order to take account of the collisional effects
more plausibly, one may replace the dispersion Eq.~(\ref{eq:20}) with

\begin{widetext}
\begin{equation}
\omega^2
\left( 1 - \Omega_1^{\prime \prime -2} \right)
\left( 1 - \Omega_2^{\prime \prime -2} \right)
- k^2
\left[
\left( 1 - \Omega_1^{\prime \prime -2} \right)
\left( 1 + \Omega_3^{\prime \prime -2} \right) +
\Omega_4^{\prime \prime -4}
\right] \approx 0,
\label{eq:48}
\end{equation}
\end{widetext}

\noindent
where, instead of Eqs.~(\ref{eq:18}) and (\ref{eq:19}), use
$\omega^{\prime \prime 2}_a = (1 + i \tilde{\nu}_{ei,a}) \omega^2$ and

\begin{widetext}
$$
\Omega_{1}^{\prime \prime -2} = \sum_{a}
{ n_{0,a} \over {\Gamma_{0,a} \omega^{\prime \prime 2}_a} },~~
\Omega_{2}^{\prime \prime -2} = \sum_{a}
{ n_{0,a} \over {\Gamma_{0,a}^3 \omega^{\prime \prime 2}_a} },
$$
\begin{equation}
\Omega_{3}^{\prime \prime -2} = \sum_{a}
{ {n_{0,a}v_{0,a}^2} \over {\Gamma_{0,a} \omega^{\prime \prime 2}_a} },~~
\Omega_{4}^{\prime \prime -2} = \sum_{a}
{ {n_{0,a}v_{0,a}} \over {\Gamma_{0,a} \omega^{\prime \prime 2}_a} },
\label{eq:49}
\end{equation}
\end{widetext}

\noindent
respectively.
For simplicity, we set the current parameters to the same as the previous ones:
$n_{0,1} = 0.1$ ($v_{0,1} = 0.9$) and $n_{0,2} = 0.9$ ($v_{0,2} = -0.1$),
but now providing the slow return current with $\Gamma_{0,2} \simeq 1$
to be resistive such as $\tilde{\nu}_{ei,2} \neq 0$
($\gg \tilde{\nu}_{ei,1}$).
Numerical calculations indicate that Eq.~(\ref{eq:48})
contains six complex solutions,
consisting of three pairs of the positive and negative solutions.
For the three unstable modes each, one can define the growth rates as
$\gamma = |\omega_i|$.

In Fig.~\ref{fig:12}, I show $\gamma$ as a function of $k$,
for the collision parameters of
$\tilde{\nu}_{ei,1} = 0$ and $\tilde{\nu}_{ei,2} = 1$, as an example.
It is found that as a whole the wave number dependence itself is
similar to that for the symmetrically counterstreaming cases
(compare Figs.~\ref{fig:7} and \ref{fig:10}).
The purely growing mode ($\tilde{\nu}_{ei,2} = 0$: crosses)
is disturbed due to the collisional effects on the slow return current
to yield the {\it n} mode, namely, the dephasing Weibel mode.
However, the dephasing effects result in only a slight increase of
the growth rate.
As shown in Fig.~\ref{fig:12} (solid curve) for $\tilde{\nu}_{ei,2} = 1$
the increasing rate is about $14\%$ at most around $k \simeq 0.5$.
I mention that for the larger collision parameter, the growth rate of
the {\it n} mode is depressed below that for the collisionless
($\tilde{\nu}_{ei,2} = 0$) case.
In addition, the dephasing effects create the unstable modes
analogous to the {\it o} and {\it p} mode that were introduced in
Sec.~\ref{sec:3b}.
The growth rate of the corresponding {\it o} mode now quite weakly depends
on the wave number.
Both the modes are found to surpass the {\it n} mode in
growing long-wavelength perturbations, as shown in the figure.
In particular, for the parameter region of
$\tilde{\nu}_{ei,2} \sim {\rm O}(1)$,
the growth rate of the {\it o} mode exceeds,
even in the short-wavelength region, that of the {\it n} mode,
that is, along the mechanism elucidated in Sec.~\ref{sec:3b},
the electron-ion collision affects the slow return current,
even if the forward beam current is in the collisionless regime.
This consequence is consistent with the previous results
obtained by carrying out the fully relativistic and
electromagnetic particle simulation for asymmetrically counterstreaming
electron currents in plasma \cite{honda00c}.

\section{\label{sec:4}Concluding remarks}

In conclusion, I have systematically investigated the details of collisional
and thermal effects on the relativistic current filamentation instability,
generalizing dispersion relation of the beam-Maxwell system.
For specific cases, the approximate dispersions have been derived
and applied to the instability analysis of typical
counterstreaming relativistic electron currents, relevant to ignitor physics.
For the symmetrically counterstreaming cases,
the particular results are summarized as follows:\\

\noindent
~~(i) Effects of electron-electron collision
suppress the relativistic Weibel instability for all wavelengths.
The effects newly create a growing wave mode,
but its growth rate is always lower than that of the suppressed
Weibel instability.

\noindent
~~(ii) Effects of electron-ion collision suppress the relativistic
Weibel instability, especially for short-wavelength perturbations.
The effects create growing oscillatory and wave mode.
For long-wavelength perturbations, the growth rates of both modes
tend to exceed the growth rate of the suppressed Weibel instability.
For the stronger collisional coupling, and lower current speed as well,
the growth rate of the oscillatory mode can, for all wavelengths,
exceed that of the suppressed Weibel instability.

\noindent
~~(iii) Effects of thermal spread simply suppress the relativistic Weibel
instability, at least, in a moderate wavelength range.\\

\noindent
For the asymmetrically counterstreaming case,
the relativistic Weibel instability for the symmetrical case is
strongly suppressed, though the electron-ion collision still affects
the slow return current, creating the unstable modes,
as mentioned above in result (ii).

The important point is that, in general,
the growing oscillatory mode is created by dephasing
a purely oscillatory mode, and the growing wave mode is created by dephasing
either a purely oscillatory wave mode or a purely growing mode.
While the collisional effects invoke phase lag, reflecting the inverse
transformation of Eqs.~(\ref{eq:13}) and (\ref{eq:18}) for electron-electron
and electron-ion collision, respectively,
the thermal effects do not dephase the purely oscillatory,
oscillatory wave, and growing mode, but involve mode dispersion.
Hopefully, intense laser-plasma interaction experiment will be able to
reproduce these fundamental consequences, though they were derived
by leaving out longitudinal modes, complexities of mode-coupling,
and so forth.

\begin{acknowledgments}
I am grateful to J.~Meyer-ter-Vehn for a useful discussion.
\end{acknowledgments}

\appendix
\section{Derivation of the generalized dispersion relation
including collisional effects and thermal corrections}

In this appendix, I briefly explain the derivation of Eq.~(\ref{eq:6}).
By linearizing the continuity Eq.~(\ref{eq:1}),
the density perturbation of electron component $a$ is described as
$n_{1,a} = n_{0,a} \Omega_k^{-1} (p_{1,a,y} / \Gamma_{0,a})$,
where $\Omega_k^{-1} = k / \omega$, and
$p_{1,a,i}$ stand for the first order quantities of
$i$-directional momentum of the component $a$.
Making use of the relations of $p_{1,a,x} = E_{1,x} / (i \omega - \nu_{ei} )$,
which are obtained by linearizing $x$ component of
the vector Eq.~(\ref{eq:2}),
the first order momenta $p_{1,a,y}$ can be expressed
as functions of the first order electric fields $E_{1,i}$.
Substituting these expressions into the linearized continuity equation
mentioned above, I obtain the first order equations for
density perturbation in the form of

\begin{widetext}
\begin{equation}
{ n_{1,a} \over n_{0,a} } =
{ \Omega_k^{-1} \over {i \omega \Gamma_{0,a}} }
{
{
v_{0,a} \Omega_k^{-1} \left[
\left( 1 - \tilde{T}_{0,\bar{a}} \Omega_k^{-2} + i \nu \right) +
i \tilde{\nu}_{ee} \upsilon_{\bar{a}a}
\right] E_{1,x}
+
\left(
1 - \tilde{T}_{0,\bar{a}} \Omega_k^{-2} + i \nu_1
\right) E_{1,y}
} \over {
\left( 1 - \tilde{T}_{0,a} \Omega_k^{-2} + i \nu \right)
\left( 1 - \tilde{T}_{0,\bar{a}} \Omega_k^{-2} + i \nu \right)
+ \tilde{\nu}_{ee}^2
}
}
\label{eq:a1}
\end{equation}
\end{widetext}

\noindent
for $\omega \neq 0$.
Here, the abbreviations of Eq.~(\ref{eq:9}) have been used.
Substituting the expressions of $p_{1,a,i}$, Eq.~(\ref{eq:a1}), and
$B_{1,z} = -\Omega_k^{-1} E_{1,x}$ derived from Eq.~(\ref{eq:3}),
into the linearized Eq.~(\ref{eq:4}),
yields the first order equations of $E_{1,i}$.
For the manipulations,
the second order terms for thermal correction of the form of
$\sim (\tilde{T}_{0,a} \Omega_k^{-2})^2$ in the products of, e.g.,
$(1-\tilde{T}_{0,1}\Omega_k^{-2}) (1-\tilde{T}_{0,2}\Omega_k^{-2})$
are neglected, such that
$(1-\tilde{T}_{0,1}\Omega_k^{-2}) (1-\tilde{T}_{0,2}\Omega_k^{-2})
\approx 1 - (\tilde{T}_{0,1} + \tilde{T}_{0,2})\Omega_k^{-2}
= 1 - T\Omega_k^{-2}$.
Finally, we arrive at the equation of the form of
${\cal D}_{j}^{~i}(k,\omega) E_{1,i} \approx 0$~($\forall E_{1,i}$),
where the determinant of the dielectric tensor is

\begin{widetext}
$$
|{\cal D}(k,\omega)| = 
\{
\left( 1 + \tau \Omega_k^{-2} \right)
\left( 1 - \Omega_k^{-2} \right) -
\left( \Omega_{21}^{-2} + i \Omega_{22}^{-2} \right)
$$
$$
-
\left[
\left( \Omega_{21,T}^{-2} + i \Omega_{22,T}^{-2} \right) +
\left( \Omega_{31}^{-2} + i \Omega_{32}^{-2} \right)
\right] \Omega_k^{-2} -
\left( \Omega_{31,T}^{-2} + i \Omega_{32,T}^{-2} \right) \Omega_k^{-4}
\}
$$
$$
\times
{
{
\left( 1 + \tau \Omega_k^{-2} \right) -
\left( \Omega_{11}^{-2} + i \Omega_{12}^{-2} \right) -
\left( \Omega_{11,T}^{-2} + i \Omega_{12,T}^{-2} \right) \Omega_k^{-2}
} \over {
\left[
\left( \Omega_{43}^{-2} + i \Omega_{44}^{-2} \right) +
\left( \Omega_{43,T}^{-2} + i \Omega_{44,T}^{-2} \right) \Omega_k^{-2}
\right] \Omega_k^{-1}
}
}
$$
\begin{equation}
-
\left[
\left( \Omega_{41}^{-2} + i \Omega_{42}^{-2} \right) +
\left( \Omega_{41,T}^{-2} + i \Omega_{42,T}^{-2} \right) \Omega_k^{-2}
\right] \Omega_k^{-1}.
\label{eq:a2}
\end{equation}
\end{widetext}

\noindent
Here, the definitions of Eq.~(\ref{eq:7}) have been used.
For $k\neq 0$, the dispersion relation can be defined by
$|{\cal D}(k,\omega)| = 0$, to give Eq.~(\ref{eq:6}).

\begin{figure*}
\resizebox{160mm}{!}{\includegraphics{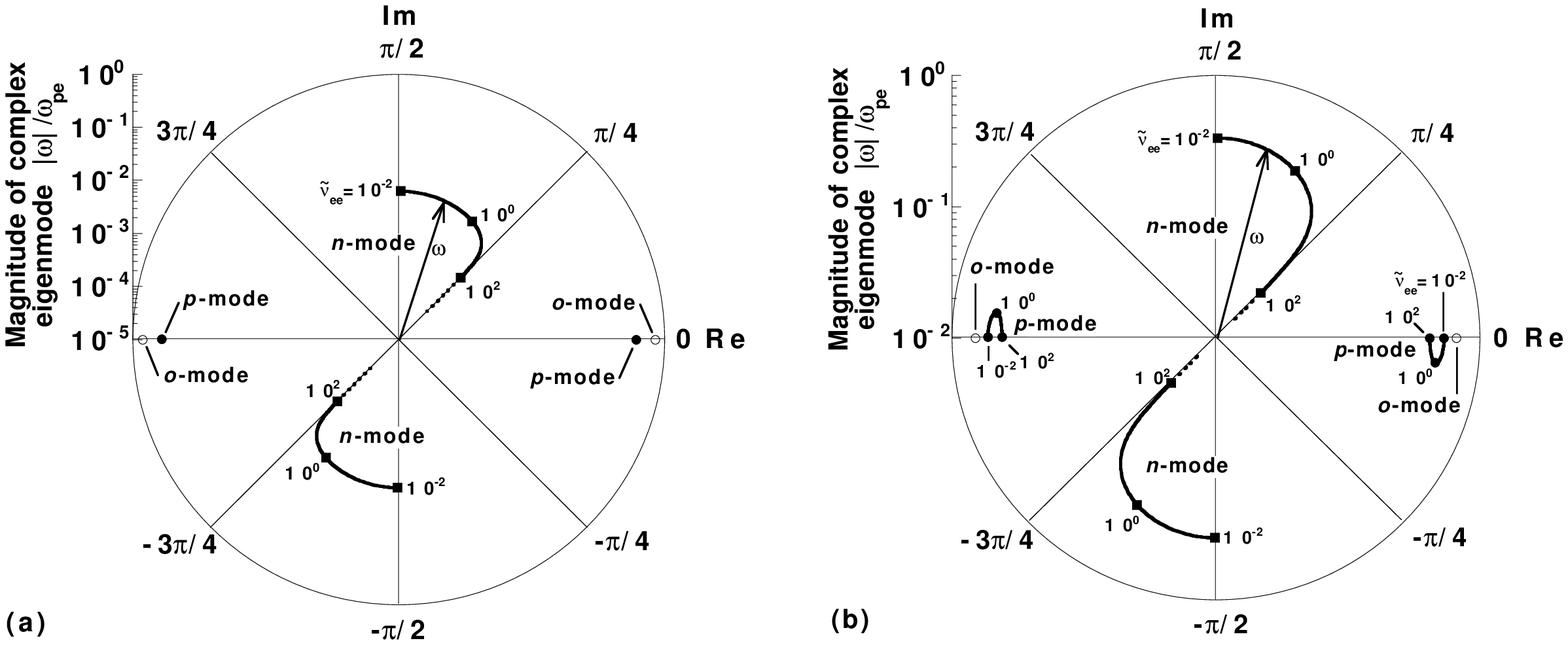}}
\caption{\label{fig:1}
Polar coordinate plots of complex eigenmodes contained in dispersion
Eq.~(\ref{eq:28}),
in the parameter range of $10^{-2} \leq \tilde{\nu}_{ee} \leq 10^2$
for (a) $k = 3.0 \times 10^{-3}$ and (b) $k = 3.0 \times 10^{-1}$:
$\omega = \pm \omega_{0p} {\rm exp} (i \theta_{0p})$
for {\it p} mode (solid curves with filled circles) and
$\omega = \pm \omega_{0n} {\rm exp} (i \theta_{0n})$
for {\it n} mode (solid curves with filled squares).
The {\it p} and {\it n} mode refer to the definition of,
in Eq.~(\ref{subeq:33a}),
$\omega_{0p} = |\omega_0(\mu=\mu_+)|$ and
$\omega_{0n} = |\omega_0(\mu=\mu_-)|$, respectively,
and in Eq.~(\ref{subeq:33b}), $\theta_{0p} = \theta_0(\mu=\mu_+)$ and
$\theta_{0n} = \theta_0(\mu=\mu_-) + \pi/2$, respectively.
For comparison, I also plot the fixed points of
$\omega = \pm \Gamma_0^{-1/2}$ for {\it o} mode (open circles).
The horizontal and vertical line crossing at the center correspond to the
real and imaginary axis, respectively.
The plots of $\omega$ can be compared to,
e.g., as indicated by arrows labeled as $\omega$,
the trajectories of arrowhead of the vectors,
whose magnitude is scaled logarithmically by the left axis.
Here, I have chosen the parameter of $v_0 = 0.9$,
setting $\tilde{\nu}_{ei} = 0$ and $\tilde{T}_{0,a} = 0$.
Note that for (a) $k = 3.0 \times 10^{-3}$,
the vectors of the {\it p} mode do not largely deviate from the real axis.
For an explanation, see the text.
}
\end{figure*}

\begin{figure*}
\resizebox{170mm}{!}{\includegraphics{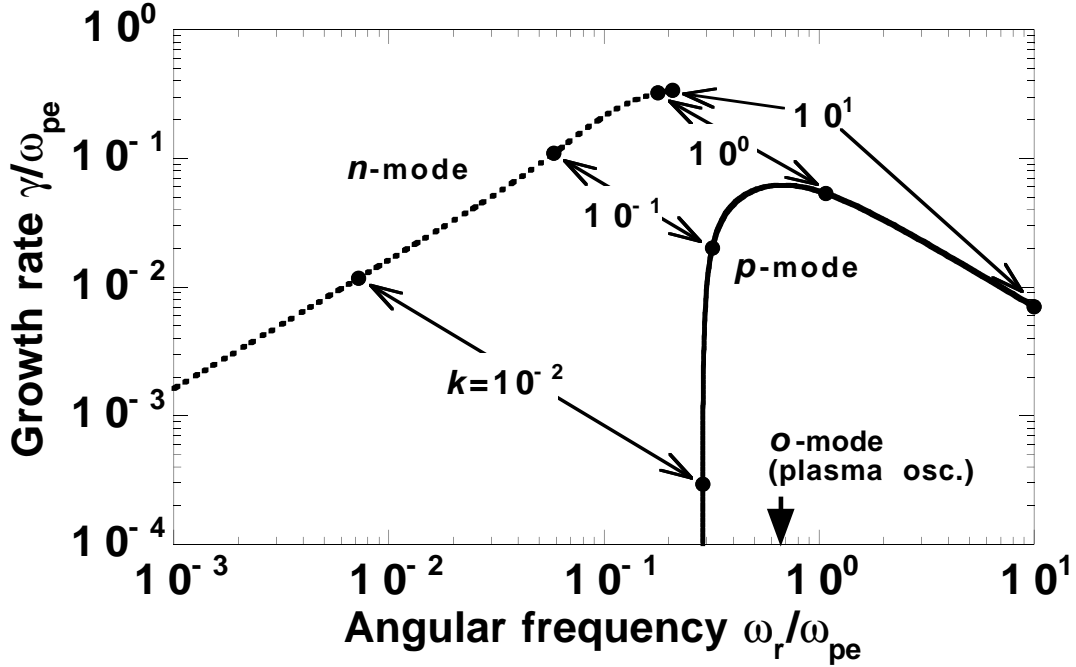}}
\caption{\label{fig:3}
The linear growth rate $\gamma$ of Eq.~(\ref{subeq:35b})
vs the angular frequency $\omega_r$ of Eq.~(\ref{subeq:35a})
for the {\it p} mode (solid curve) and the {\it n} mode (dotted curve),
varying the wave number $k$ as a parameter.
Here, I have chosen the parameters of $v_0 = 0.9$ and $\tilde{\nu}_{ee} = 1$,
setting $\tilde{\nu}_{ei} = 0$ and $\tilde{T}_{0,a} = 0$.
For comparison, the angular frequency of the {\it o} mode given by
Eq.~(\ref{eq:30}) is also indicated by a bold arrow.
}
\end{figure*}

\begin{figure*}
\resizebox{160mm}{!}{\includegraphics{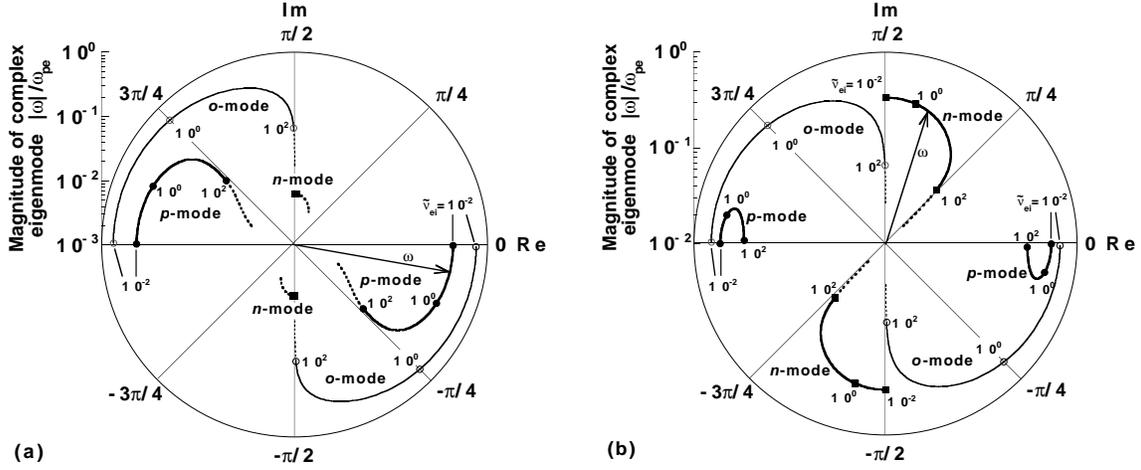}}
\caption{\label{fig:6}
Polar coordinate plots of complex eigenmodes contained in dispersion
Eq.~(\ref{eq:36}),
in the parameter range of $10^{-2} \leq \tilde{\nu}_{ei} \leq 10^2$
for (a) $k = 3.0 \times 10^{-3}$ and (b) $k = 3.0 \times 10^{-1}$:
$\omega = \pm \omega_{0p} {\rm exp} (i \theta_{0p})$
for {\it p} mode (solid curves with filled circles) and
$\omega = \pm \omega_{0n} {\rm exp} (i \theta_{0n})$
for {\it n} mode (solid curves with filled squares), and
Eq.~(\ref{eq:37}) for {\it o} mode independent of $k$
(hair solid curves with open circles).
The {\it p} and {\it n} mode refer to Eq.~(\ref{eq:38})
for $\mu = \mu_+$ and $\mu_-$, respectively, and
the definitions of $\omega_{0p}$, $\omega_{0n}$, $\theta_{0p}$, and
$\theta_{0n}$ are the same as those in Fig.~\ref{fig:1} caption.
The horizontal and vertical line correspond to the
real and imaginary axis, respectively.
The plots of $\omega$ can be compared to the trajectories of arrowhead of
the vectors, whose magnitude is scaled by the left axis.
Here, I have chosen the parameter of $v_0 = 0.9$,
setting $\tilde{\nu}_{ee} = 0$ and $\tilde{T}_{0,a} = 0$.
Note that for (a) $k = 3.0 \times 10^{-3}$,
the vectors of the {\it n} mode do not largely deviate from the imaginary axis
for the parameter range being considered (see text).
}
\end{figure*}

\begin{figure*}
\resizebox{170mm}{!}{\includegraphics{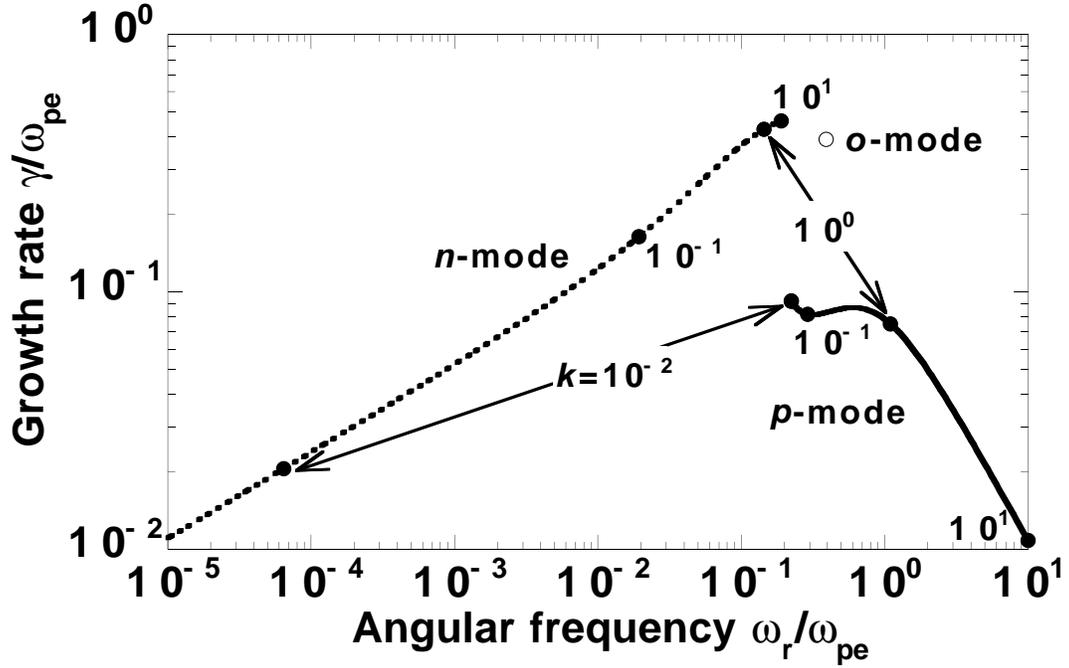}}
\caption{\label{fig:8}
The linear growth rate $\gamma$ of Eq.~(\ref{subeq:41b})
vs the angular frequency $\omega_r$ of Eq.~(\ref{subeq:41a})
for the {\it p} mode (solid curve) and the {\it n} mode (dotted curve),
varying the wave number $k$ as a parameter.
An open circle indicates the point given by Eq.~(\ref{eq:37})
for the {\it o} mode.
Here, I have chosen the parameters of $v_0 = 0.9$ and $\tilde{\nu}_{ei} = 1$,
setting $\tilde{\nu}_{ee} = 0$ and $\tilde{T}_{0,a} = 0$.
}
\end{figure*}

As it is well known, if the imaginary part of the complex eigenvalues
$\omega$ is much smaller than the real part,
one can calculate them by the Taylor expansion method \cite{dendy90}.
However, this is not the case being considered,
as seen in Figs.~\ref{fig:1}, \ref{fig:3}, \ref{fig:6}, and \ref{fig:8}.
That is why, I have attempted to directly solve the complex Eq.~(\ref{eq:6})
to extract the eigenvalues.


\end{document}